\documentclass[preprintnumbers,secnumarabic,twocolumn]{revtex4}
\pdfoutput=1
\usepackage{graphicx}
\usepackage{amsmath,amssymb,mathrsfs}
\usepackage{fancyhdr}
\usepackage{psfrag}
\usepackage{array,bbm} 
\usepackage{url}

\newcommand*{\ewgroup}{\ensuremath{SU(2)_L \times U(1)_Y}}
\newcommand*{\CPs}{\ensuremath{\mathrm{CP_s}}}
\newcommand*{\CP}{\ensuremath{\text{CP}}}
\newcommand*{\CPg}{\ensuremath{\text{CP}_g}}
\newcommand*{\CPi}{\ensuremath{\mathrm{CP}_g^{(i)}}}
\newcommand*{\CPii}{\ensuremath{\text{CP}_g^{(ii)}}}
\newcommand*{\CPa}{\ensuremath{\text{CP}_{g,1}^{(ii)}}}
\newcommand*{\CPb}{\ensuremath{\text{CP}_{g,2}^{(ii)}}}
\newcommand*{\CPc}{\ensuremath{\text{CP}_{g,3}^{(ii)}}}
\newcommand*{\trans}{\mathrm{T}}                     % transposed
\newcommand*{\unitmatrix}{\mathbbm{1}}
\newcommand*{\tvec}[1]{\ensuremath{\boldsymbol{\mathrm{#1}}}}           % 3 vec
\newcommand*{\tmat}[1]{\underline{#1}}             % 2x2 matrix

\makeatletter       % Macros for arabic section numbering

\renewcommand{\p@subsection}{}

\makeatother

\DeclareMathOperator{\re}{Re}
\DeclareMathOperator{\im}{Im}

\begin{document}

\preprint{HD-THEP-08-29}

\title{On the phenomenology of a two-Higgs-doublet model with
maximal CP symmetry at the LHC}

% EPJC
%\author{M. Maniatis, {\email{M.Maniatis@thphys.uni-heidelberg.de}},\\
%O. Nachtmann, {\email{O.Nachtmann@thphys.uni-heidelberg.de}}
%    	}
%Revtex
\author{M. Maniatis}
    \email[E-mail: ]{M.Maniatis@thphys.uni-heidelberg.de}
\author{O. Nachtmann}
    \email[E-mail: ]{O.Nachtmann@thphys.uni-heidelberg.de}

% EPJC
%\institute{
%Institut f\"ur Theoretische Physik, Philosophenweg 16, D-69120 Heidelberg, Germany 
%}

% revtex
\affiliation{
Institut f\"ur Theoretische Physik, Philosophenweg 16, 69120
Heidelberg, Germany
}

%\abstract{
\begin{abstract}
Predictions for LHC physics are worked out for a two-Higgs-doublet model
having four generalized CP symmetries. In this
{\em maximally-CP-symmetric model}
(MCPM) the first fermion family is, at tree level, uncoupled to the Higgs fields and
thus massless. The second and third fermion families have a very symmetric coupling
to the Higgs fields. But through the electroweak symmetry breaking a large mass
hierarchy is generated between these fermion families. Thus, the fermion mass spectrum
of the model presents a rough approximation to what is observed in Nature.
In the MCPM there are, as in every two-Higgs-doublet model,
five physical Higgs bosons, three neutral ones and a charged pair. 
In the MCPM the couplings of the Higgs bosons to the fermions
are completely fixed. This allows us to present clear predictions
for the production at the LHC and for the decays of the physical Higgs bosons.
As salient feature we find rather large cross sections for Higgs-boson
production via Drell--Yan type processes. With experiments at the LHC
it should be possible to check these predictions.
%}
\end{abstract}

\maketitle

%%%%%%%%%%%%%%%%%%%%%%%%%%%%%%%%%%%%%%%%%%%%%%%%%%%%%%%%%%%%%%%%%%%%%%%%
% Introduction
%
%%%%%%%%%%%%%%%%%%%%%%%%%%%%%%%%%%%%%%%%%%%%%%%%%%%%%%%%%%%%%%%%%%%%%%%
\section{Introduction}
\label{sec-intro}

The Standard Model (SM) of particle physics is very successful
in describing the currently known experimental data;
see \cite{Amsler:2008zz} for a review.
Nevertheless, the SM leaves open a number of theoretical questions.
Thus, various extensions of the SM have been studied extensively.
With the start-up of the LHC we can hope that
experiments will soon give decisive answers in which way
- if at all - the SM has to be extended; see \cite{Ellis:2007vz}
for a brief overview of these topics.

In this paper we shall study a particular two-Higgs-doublet model (THDM)
and develop its LHC phenomenology.
The model, which we want to call
maximally\--CP\--sym\-metric model (MCPM) 
for reasons which will become clear later, has the field content as in the SM except for the
Higgs sector, where we have two Higgs doublets instead of only one.
Many versions of THDMs have been studied in the literature; see 
\cite{Kobayashi:1973fv,Gunion:1989we,Cvetic:1993cy,Ginzburg:2004vp,Gunion:2005ja,Barbieri:2005kf,Branco:2005em,Nishi:2006tg,Barbieri:2006dq,Ivanov:2006yq,Fromme:2006cm,Ivanov:2007de,Barroso:2007rr,Gerard:2007kn}
and references therein.
In our group we have studied various aspects of the most general
THDM in \cite{Maniatis:2006fs,Maniatis:2007vn}. A class of
interesting models having a maximal number of generalized
CP symmetries was found. In \cite{Maniatis:2007de} these
models were studied in detail and it was shown that the
requirement of maximal CP invariance led to a very interesting
structure for the coupling of fermions to the Higgs fields.
Maximal CP invariance requires more than one fermion family
if fermions are to get non-zero masses.
With the additional requirement of absence of flavor-changing
neutral currents at tree level and of mass-degenerate massive
fermions a unique Lagrangian was derived. 
This Lagrangian is very symmetric between the second and third
fermion families {\em before} electroweak symmetry breaking
(EWSB) occurs. But after EWSB the third family becomes massive,
the second family stays massless. In this model also the
first family is massless and the Cabibbo--Kobayashi--Maskawa (CKM)
matrix equals the unit matrix.
Of course, all this is not exactly as observed in Nature. 
But, on the other hand, it may be a starting point to
understand some aspects of the large fermion mass hierarchies
observed experimentally. 

In the present paper we shall work out concrete predictions for
LHC physics which follow from the two-Higgs-doublet model with maximal CP invariance,
the MCPM,
having the large fermion mass hierarchies as discussed in \cite{Maniatis:2007de}.
In Sect.~\ref{secII} we recall the main
features of the Lagrangian. In Sect.~\ref{secIII} we give
our predictions for the decays of the physical Higgs particles
of the MCPM. Section~\ref{secIV} deals with Higgs-boson production
at the LHC. We draw our conclusions in Sect.~\ref{secC}.
In Appendix~\ref{appA} we give the explicit form of the Lagrangian
and some Feynman rules of the MCPM. If the MCPM, in
the strict symmetry limit, represents not too bad an
approximation to the real world then this should also be
true for its LHC phenomenology as discussed in this paper.
Thus, our work should be considered as presenting the {\em generic
features} of this phenomenology.

%%%%%%%%%%%%%%%%%%%%%%%%%%%%%%%%%%%%%%%%%%%%%%%%%%%%%%%%%%%%%%%%%%%%%%%%
% Section II, the model
%
%%%%%%%%%%%%%%%%%%%%%%%%%%%%%%%%%%%%%%%%%%%%%%%%%%%%%%%%%%%%%%%%%%%%%%%
\section{The model}
\label{secII}

% gauge invariant functions
A detailed study of the MCPM can be found in Ref.~\cite{Maniatis:2007de}.
Here we want to recall the motivation and the essential steps to
construct this model.

The general gauge-invariant and renormalizable potential $V(\varphi_1, \varphi_2)$
of the two Higgs doublets $\varphi_1$ and $\varphi_2$ is a hermitian linear
combination of the terms
\begin{equation}
\varphi_i^\dagger \varphi_j,\quad
(\varphi_i^\dagger \varphi_j)
(\varphi_k^\dagger \varphi_l)\;, 
\end{equation}
with $i,j,k,l \in \{1,2\}$.
The \ewgroup~invariant scalar products are arranged into the hermitian, positive semi definite, 
$2 \times 2$ matrix
\begin{equation}
\label{eq-kmat}
\tmat{K}(x) :=
\begin{pmatrix}
  \varphi_1^{\dagger}\varphi_1 & \varphi_2^{\dagger}\varphi_1 \\
  \varphi_1^{\dagger}\varphi_2 & \varphi_2^{\dagger}\varphi_2
\end{pmatrix}\,.
\end{equation}
Its decomposition reads
\begin{equation}\label{2.4}
\tmat{K}(x)
 = \frac{1}{2}\left( K_0(x)\unitmatrix_2
      + \tvec{K}(x)\,\tvec{\sigma} \right)
\end{equation}
with Pauli matrices $\sigma^a\ (a=1,2,3)$. In this way one defines
the real {\em gauge-invariant functions}
\begin{equation}
\label{eq-kdef}
\begin{aligned}
K_0 &= \varphi_1^{\dagger} \varphi_1 + \varphi_2^{\dagger} \varphi_2, &
K_1 &= 2 \re \varphi_1^\dagger \varphi_2,\\
K_3 &= \varphi_1^{\dagger} \varphi_1 - \varphi_2^{\dagger} \varphi_2, &
K_2 &= 2 \im \varphi_1^\dagger \varphi_2\,.\\
\end{aligned}
\end{equation}
In terms of these functions the general THDM potential can be 
written in the simple form
\begin{align}\label{2.8}
V =
  &\; \xi_0\, K_0(x) + \tvec{\xi}^\trans\, \tvec{K}(x)
  + \eta_{00}\, K_0^2(x)\notag\\ 
  &+ 2\, K_0(x)\,\tvec{\eta}^\trans\, \tvec{K}(x)
  + \tvec{K}^\trans(x)\, E\, \tvec{K}(x)\,,
\end{align}
with $\tvec{K}=(K_1, K_2, K_3)^\trans$ and parameters
$\xi_0$, $\eta_{00}$, three-component vectors $\tvec{\xi}$, $\tvec{\eta}$
and the $3\times 3$ matrix $E=E^\trans$.
All parameters in~\eqref{2.8} are real. 

% CP transformations
One now proceeds to study CP transformations in the general THDM.
Writing the Higgs potential in the form~\eqref{2.8} 
one finds a simple geometric picture for CP transformations.
The {\em standard} CP transformation (\CPs) of the Higgs doublets
is
\begin{equation}
\label{eq-simCP}
\varphi_i(x) \xrightarrow{\CPs} \phantom{-}\varphi_i^*(x')\,, \qquad (i=1,2)\,,
\end{equation}
where, due to the parity transformation $x'=(x_0, -\tvec{x})^\trans$.
In terms of the gauge invariant functions, this \CPs~transformation is simply
$K_0(x)\rightarrow K_0(x')$ and
\begin{equation}
\label{eq-simCPK}
\begin{split}
K_1(x) &\rightarrow \phantom{-} K_1(x')\;,\\
K_2(x) &\rightarrow - K_2(x')\;,\\
K_3(x) &\rightarrow \phantom{-} K_3(x')\;.
\end{split}
\end{equation}
Geometrically, this is a reflection on the 1--3 plane in 
$\tvec{K}$ space in addition to the argument change. Motivated
by this geometric picture,
{\em generalized} CP transformations (\CPg) corresponding to
reflections on planes ($\CPii$)  as well as 
to the point reflection ($\CPi$) in $\tvec{K}$ space were
studied in~\cite{Maniatis:2007vn,Maniatis:2007de}.
The $\CPi$ transformation is
given by $K_0(x)\rightarrow K_0(x')$ and
\begin{equation}
\label{eq-point}
\tvec{K}(x) \rightarrow - \tvec{K}(x')
\end{equation}
and plays a central role in the construction of
the MCPM. In~\cite{Maniatis:2007de} some distinguishing features
of this transformation are discussed.
In terms of the original Higgs doublets these \CPg~transformations 
read generically
\begin{equation}
\label{eq-W}
\varphi_i(x)  \rightarrow W_{ij} \varphi^\ast_j(x')\,.
\end{equation}
The $2\times 2$ matrices $W$ corresponding
to the transformations $\CPi$ and to $\text{CP}_{g,a}^{(ii)}$
($a=1,2,3$), the reflections on the coordinate 
planes in $\tvec{K}$ space,
are given in
the second row of Tab.~\ref{tab2}, where we defined
\begin{equation}
\epsilon=
\begin{pmatrix}
\phantom{+}0 & \phantom{+}1\\
-1 & \phantom{+}0
\end{pmatrix}\;.
\end{equation}
The transformation $\CPb$ is, of course,
just $\CPs$ given in \eqref{eq-simCP}, \eqref{eq-simCPK}.
For $\CPa$ ($\CPc$) the transformation
of the $\tvec{K}$ vector is similar to~\eqref{eq-simCPK}
but with the sign change for $K_1$ ($K_3$).
\begin{table}
\centering
\begin{tabular}{ll|c|c|c}
\;\;\; \CPg \;\;\;& & $\;\;W\;\;$ & $\;\;U_R\;\;$ & $\;\;U_L\;\;$ \\
\hline
point reflection & $\CPi$ &$\epsilon$ & $\phantom{+}\epsilon$ & $\sigma^1$\\
2--3 plane reflection & $\CPa$ &$\sigma^3$ & $-\sigma^3$ & $\unitmatrix_2$\\
1--3 plane reflection & $\CPb$ &$\unitmatrix_2$ & $\phantom{+}\unitmatrix_2$ & $\unitmatrix_2$\\
1--2 plane reflection & $\CPc$ &$\sigma^1$ & $-\sigma^1$ & $\sigma^1$\\
\end{tabular}
\caption{\label{tab2}The matrices $W$ of (\ref{eq-W})
and $U_L$ and $U_R$ of (\ref{eq15}) for the
four generalized \CP\ invariances.}
\end{table}
%

% MCPM construction
Now we are in a position to recall the 
construction principles of the MCPM,
that is, a THDM which respects all generalized
\CP~symmetries of Tab.~\ref{tab2}.
We start with the THDM Higgs potential~\eqref{2.8}. 
Requiring it to be symmetric under the generalized
\CP~transformation $\CPi$ leads 
with a suitable basis choice to
\begin{equation}
\label{eq8}
V(\varphi_1, \varphi_2) =
\xi_0 K_0 + \eta_{00} K_0^2
+ \mu_1 K_1^2 + \mu_2 K_2^2 + \mu_3 K_3^2\;.
\end{equation}
Note that here $K_a$, ($a=1,2,3$) enter
only quadratically. This implies that
the potential $V$ of \eqref{eq8} is also invariant under
the transformation $\CPs \equiv \CPb$ which
just changes the sign of the component $K_2$; see \eqref{eq-simCPK}.
Similarly one finds invariance
of $V$~\eqref{eq8} under $\CPa$
and $\CPc$. Thus, the potential is
invariant under the point reflection symmetry~\eqref{eq-point}
as well as all three different reflections on the coordinate 
planes in $\tvec{K}$ space. In this way
the Higgs potential of the MCPM is determined.

The next step is to extend these \CPg~symmetries to the Yukawa terms,
which couple the fermions $\psi(x)$ to the Higgs doublets.
We define the generalized \CP~transformations of the fermions 
generically as
\begin{alignat}{2}
\label{eq15}
\CPg\,:\quad
&&\psi_{\alpha\,L}(x) &\rightarrow
   U_{L\,\alpha\beta}\,\gamma^0\,S(C)\,
   \bar{\psi}_{\beta\,L}^\trans(x')\,,
\notag\\
&&\psi_{\alpha\,R}(x) &\rightarrow
   U_{R\,\alpha\beta}\,\gamma^0\,S(C)\,
   \bar{\psi}_{\beta\,R}^\trans(x')
\end{alignat}
with family indices $\alpha$, $\beta$, $S(C) = i \gamma^2 \gamma^0$ the
usual matrix of charge conjugation, and unitary
matrices $U_L$ and $U_R$.
As shown in the detailed study~\cite{Maniatis:2007de}
having only one family coupled to the Higgs bosons
in a $\CPi$-symmetric way
leads necessarily to vanishing Yukawa couplings, that is,
to massless fermions.
Thus, in the MCPM two families are coupled  
via Yukawa terms to the Higgs doublets.
By convention these families are given the indices two and three.
One finds that the Yukawa interactions are
highly restricted
requiring them to be invariant under 
the generalized \CP~transformations
of Tab.~\ref{tab2} for 
the fermions~\eqref{eq15} and Higgs doublets~\eqref{eq-W}. 
Moreover, the Yukawa couplings are
uniquely defined, if in addition to
these \CPg~invariances one requires
non-degenerate fermion masses
and absence
of large flavor-changing neutral currents (FCNCs).
The corresponding matrices $U_L$ and $U_R$ are
presented in the last two rows in Tab.~\ref{tab2}.
Eventually, one
ends up with the Yukawa part of the Lagrangian
of the MCPM in the form
\begin{align}
\label{eq11}
\mathscr{L}_{\mathrm{Yuk}}(x) = 
  -c^{(1)}_{l\,3} & \;\Bigg\{
    \bar{\tau}_{R}(x)\,\varphi_1^\dagger(x)
    \begin{pmatrix} \nu_{\tau\,L}(x) \\ \tau_{L}(x) \end{pmatrix}
\notag\\ &
    -\bar{\mu}_{R}(x)\,\varphi_2^\dagger(x)
    \begin{pmatrix} \nu_{\mu\,L}(x) \\ \mu_{L}(x) \end{pmatrix}
    \Bigg\}
\notag\\
  +c^{(1)}_{u\,3} &\;\Bigg\{
    \bar{t}_{R}(x)\,\varphi_1^\trans(x)\,\epsilon
    \begin{pmatrix} t_{L}(x) \\ b_{L}(x) \end{pmatrix}
\notag\\ &
    -\bar{c}_{R}(x)\,\varphi_2^\trans(x)\,\epsilon
    \begin{pmatrix} c_{L}(x) \\ s_{L}(x) \end{pmatrix}
    \Bigg\}
\notag\\
  -c^{(1)}_{d\,3} &\;\Bigg\{
    \bar{b}_{R}(x)\,\varphi_1^\dagger(x)
    \begin{pmatrix} t_{L}(x) \\ b_{L}(x) \end{pmatrix}
\notag\\ &
    -\bar{s}_{R}(x)\,\varphi_2^\dagger(x)
    \begin{pmatrix} c_{L}(x) \\ s_{L}(x) \end{pmatrix}
    \Bigg\}
		+ h.c.
\end{align}
where $c^{(1)}_{l\,3}$, $c^{(1)}_{u\,3}$ and $c^{(1)}_{d\,3}$ 
are real positive constants, determined by the fermion masses
as discussed below. The first family remains uncoupled 
-- at tree level -- to the Higgs bosons in the MCPM.

Now we come to the questions of stability and EWSB in the MCPM. As discussed in
\cite{Maniatis:2007vn,Maniatis:2007de} the MCPM is stable,
produces the correct breaking $\ewgroup \rightarrow U(1)_{\mathrm{em}}$,
and has no zero mass or mass degenerate Higgs bosons
if and only if the parameters of $V$ in~\eqref{eq8} satisfy
\begin{gather}
%\begin{split}
\mu_1 > \mu_2 > \mu_3\;,\nonumber\\
\eta_{00} >0 \;,\nonumber\\
\label{eq10}
\mu_a + \eta_{00} > 0, \quad {\text{for  }} a=1,2,3 \;,\\
\xi_0 <0\;,\nonumber\\
\mu_3 <0\nonumber\;.
%\end{split}
\end{gather}
Through EWSB
only the Higgs doublet $\varphi_1$ gets a vacuum expectation value
(VEV). In the unitary gauge we have
\begin{equation}
\label{eq17}
\varphi_1(x) =
\frac{1}{\sqrt{2}}
\begin{pmatrix}
0 \\ v_0 + \rho'(x)
\end{pmatrix}\;,
\end{equation}
\begin{equation}
\label{eq18}
\varphi_2(x) =
\begin{pmatrix}
H^+(x) \\ 
\frac{1}{\sqrt{2}} ( h'(x) + i h''(x) )
\end{pmatrix}\;,
\end{equation}
where $\rho'(x)$, $h'(x)$ and $h''(x)$ are the 
real fields corresponding to the physical neutral Higgs 
particles. The fields $H^+(x)$ and 
$H^-(x) = \big( H^+(x)\big)^\ast$
correspond to the physical charged Higgs pair.
In \eqref{eq17} $v_0$ is the standard VEV
\begin{equation}
\label{eq19}
v_0 \approx 246 \text{ GeV}\;,
\end{equation}
which is given in terms of the original potential
parameters of \eqref{eq8} by
\begin{equation}
\label{eq20}
v_0 = \sqrt{ \frac{ -\xi_0 }{ \eta_{00} + \mu_3 }} \;.
\end{equation}
Inserting~\eqref{eq17} and \eqref{eq18}
into~\eqref{eq-kdef} and
\eqref{eq8} it is straightforward to
calculate
the masses of the physical Higgs fields
in terms of the original parameters
\begin{equation}
\label{eq21}
\begin{split}
m_{\rho'}^2\;\;&= 2 (- \xi_0)\;,\\
m_{h'}^2\;\;&= 2 v_0^2 (\mu_1 - \mu_3)\;,\\
m_{h''}^2\;&= 2 v_0^2 (\mu_2 - \mu_3)\;,\\
m_{H^\pm}^2&= 2 v_0^2 (- \mu_3)\;.
\end{split}
\end{equation}
Conversely, one can express the original parameters
$\xi_0,...,\mu_3$ by $v_0$ and the Higgs-boson masses
\begin{equation}
\label{eq22}
\begin{split}
\xi_0 &= -\frac{1}{2} m_{\rho'}^2\;,\\
\eta_{00} &= \frac{1}{2 v_0^2} (m_{H^\pm}^2+m_{\rho'}^2)\;,\\
\mu_1 &= \frac{1}{2 v_0^2} (m_{h'}^2-m_{H^\pm}^2)\;,\\
\mu_2 &= \frac{1}{2 v_0^2} (m_{h''}^2-m_{H^\pm}^2)\;,\\
\mu_3 &= -\frac{1}{2 v_0^2} m_{H^\pm}^2\;.
\end{split}
\end{equation}
The stability
and correct \ewgroup~symmetry breaking 
conditions~\eqref{eq10}
require positive squared masses and
%The conditions (\ref{eq10}) are always satisfied for 
%positive squared masses and
\begin{equation}
\label{eq23}
m_{h'}^2 >m_{h''}^2\;.
\end{equation}
Thus, \eqref{eq23} is the only strict relation
for the Higgs-boson masses which one gets in the MCPM.
On the other hand, if we require that the
Higgs sector has weak couplings only, we
should have $\eta_{00}$, $|\mu_1|$,
$|\mu_2|$ and $|\mu_3|$ to be
less than or equal to
a number of ${\cal O}(1)$.
From \eqref{eq21} we expect
then that the masses
of $h'$, $h''$ and $H^\pm$ should
be less than about $2 v_0 \approx 500$~GeV.
But by no means should this be considered
as a necessary upper bound for the
Higgs-boson masses in the MCPM.

Upon EWSB the Yukawa term (\ref{eq11})
produces masses for the charged fermions of
the third family. Inserting~\eqref{eq17} and
\eqref{eq18} into \eqref{eq11} gives 
%(see (121) of \cite{Maniatis:2007de})
\begin{equation}
\label{eq24}
\begin{split}
m_\tau &= c^{(1)}_{l\,3} \frac{v_0}{\sqrt{2}}\;,\\
m_t &= c^{(1)}_{u\,3} \frac{v_0}{\sqrt{2}}\;,\\
m_b &= c^{(1)}_{d\,3} \frac{v_0}{\sqrt{2}}\;.
\end{split}
\end{equation}

The fermions of the second and the first families stay
massless in the fully-symmetric theory at tree level.
Of course, this is only an approximation
valid for the tree-level investigations. 
Fortunately, from
the numerical studies which follow below, we will 
see that the main features of the LHC phenomenology
of the MCPM are insensitive to the first- and
second-family masses due to their smallness.

The next task is to express the Lagrangian
of the MCPM in terms of the physical fields in
the unitary gauge. This is done in Appendix \ref{appA}.
From there the Feynman rules of the MCPM can be read off.
In Appendix \ref{appA} we give these
rules for the three-point vertices which
are relevant for us in the following. Some
salient features are as follows.

\begin{itemize}
\item The neutral Higgs boson $\rho'$ couples to 
the third-family fermions as the physical Higgs boson $\rho'_{SM}$
of the SM.
\item The neutral Higgs boson $h'$ has a scalar
coupling to the {\em second}-family fermions.
The Higgs boson $h''$ which is lighter than $h'$
has a pseudoscalar coupling to the {\em second}-family
fermions.
But the coupling constants for $h'$ and $h''$ are 
proportional to the masses of the {\em third}-family fermions, 
that is, to $m_{\tau}$, $m_t$ and $m_b$.
\item Also the charged Higgs bosons $H^\pm$ couple
only to the second-family fermions but again with
coupling constants proportional to the masses
of the third-family fermions.
\end{itemize}

As we shall see in the following these
features lead to quite distinct
phenomenological predictions
of the MCPM for LHC physics. 

We summarize this section. We have recalled the
construction principles of the model which has
the four generalized \CP~symmetries of Tab.~\ref{tab2}.
As can easily be seen from~\eqref{eq8}
this is the maximal number of such symmetries,
including $\CPi$, one can have in a THDM 
if one requires absence of zero mass and
mass degenerate physical Higgs bosons. Thus,
the name {\em maximally--\CP--symmetric model}, MCPM,
seems justified. The extension of the four generalized
\CP~symmetries to the Yukawa interaction gave 
drastic restrictions for the family structure
of the model and led, finally, with some
additional arguments to the coupling~\eqref{eq11}.
The remaining sections of this paper are devoted
to discussing physical consequences of the MCPM.

%%%%%%%%%%%%%%%%%%%%%%%%%%%%%%%%%%%%%%%%%%%%%%%%%%%%%%%%%%%%%%%%%%%%%%%%
% section III
%
%%%%%%%%%%%%%%%%%%%%%%%%%%%%%%%%%%%%%%%%%%%%%%%%%%%%%%%%%%%%%%%%%%%%%%%
\section{Higgs-boson decays}
\label{secIII}

The decays of the Higgs particles of the MCPM
which are possible at tree level can be directly read off
from the Lagrangian in the form given in (\ref{eqA2})
of Appendix \ref{appA}.
We have decays of a Higgs particle into a fermion and
an antifermion, and of a Higgs particle into another
Higgs particle plus a gauge boson $W$ or $Z$.
Furthermore,
we could have decays of one Higgs boson into two other Higgs bosons
and one Higgs boson into another Higgs boson plus two gauge bosons
if the mass differences of the various Higgs bosons are large
enough. In the following we shall restrict ourselves to discussing
the tree-level results for
the fermionic and the Higgs boson plus gauge boson decays and
the results for the loop-induced two-photon and two-gluon decays. 

%%%%%%%%%%%%%%%%%%%%%%%%%%%%%%%%%%%%%%%%%%%%%%%%%%%%%%%%%%%%%%%%%%%%%%%%
% subsection III.1
%
%%%%%%%%%%%%%%%%%%%%%%%%%%%%%%%%%%%%%%%%%%%%%%%%%%%%%%%%%%%%%%%%%%%%%%%
\subsection{Fermionic decays}
\label{secIII1}

\begin{figure}[t] 
\begin{tabular}{m{0.5\linewidth}m{0.4\linewidth}}
\includegraphics[width=.7\linewidth]{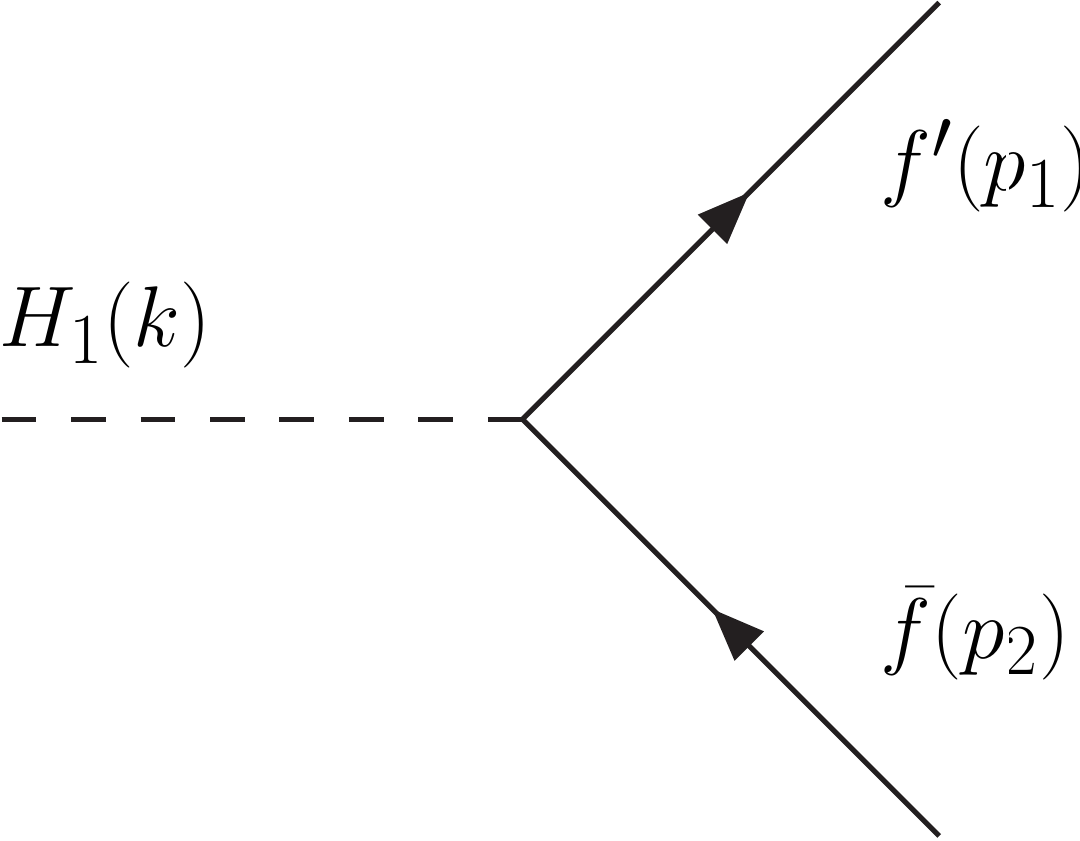} 
\hspace{0.1\linewidth}
&
{\large $-i  \frac{1}{v_0} (a + b \;\gamma_5)$}
\\
\end{tabular}\\
\caption{\label{fig1}
The diagram for the generic decay $H_1 \rightarrow f' \bar{f}$
and the corresponding analytic expression for the vertex.}
\end{figure}

The generic fermionic decay of a Higgs particle $H_1$
is
\begin{equation}
\label{eq25}
H_1(k) \rightarrow f'(p_1) + \bar{f}(p_2)
\end{equation}
where $f$ and $f'$ denote the fermions and the momenta
are indicated in brackets. The corresponding diagram and analytic
expression at tree level for the vertex are shown in Fig.~\ref{fig1}.
The possible decays together with the corresponding coupling constants
$a$ and $b$ are listed in Tab. \ref{tab3}. There, $N_c^f$ is the
color factor which equals $1$ for leptons and $3$ for quarks. The decay 
rate for the generic decay (\ref{eq25}) is calculated as
\begin{multline}
\label{eq26}
\Gamma (H_1 \rightarrow f' + \bar{f}) =\\
\frac{N_c^f}{8 \pi v_0^2}
\frac{w(m_{H_1}^2, m_f^2, m_{f'}^2)}{m_{H_1}^2}\;
m_{H_1} \; \theta(m_{H_1}-m_f-m_{f'})\\
\bigg\{
	|a|^2 + |b|^2 
	-\frac{(m_f + m_{f'})^2}{m_{H_1}^2} |a|^2
	-\frac{(m_f - m_{f'})^2}{m_{H_1}^2} |b|^2
\bigg\}\;.
\end{multline}
\begin{table*}
\centering
\begin{tabular}{c||c|c|c|c|c|c}
$H_1$ & $f'$ & $\bar{f}$ & $a$ & $b$ & $|a|^2+|b|^2$ & $N_c^f$\\
\hline
$\rho'$ & $\tau$ & $\bar{\tau}$ & \phantom{+}$m_{\tau}$ & 0 & $m_\tau^2$ & $1$\\
	    & $t$      & $\bar{t}$    & \phantom{+}$m_t$      & 0 & $m_t^2$    & $3$\\
	    & $b$      & $\bar{b}$    & \phantom{+}$m_b$      & 0 & $m_b^2$    & $3$\\
\hline
$h'$    & $\mu$  & $\bar{\mu}$  & $-m_{\tau}$           & 0 & $m_\tau^2$ & $1$\\
        & $c$      & $\bar{c}$    & $-m_t$                & 0 & $m_t^2$    & $3$\\
        & $s$      & $\bar{s}$    & $-m_b$                & 0 & $m_b^2$    & $3$\\
\hline
$h''$   & $\mu$  & $\bar{\mu}$  & $0$   & $-i m_{\tau}$ & $m_\tau^2$ & $1$\\
        & $c$      & $\bar{c}$    & $0$   & $\phantom{+}i m_t$      & $m_t^2$    & $3$\\
        & $s$      & $\bar{s}$    & $0$   & $-im_b$       & $m_b^2$    & $3$\\
\hline
$H^+$   & $\nu_\mu$ & $\bar{\mu}$ & $-m_\tau/\sqrt{2}$    & $-m_\tau/\sqrt{2} $ & $m_\tau^2$ & $1$\\
        & $c$       & $\bar{s}$   & $(m_t-m_b)/\sqrt{2} $ & $-(m_t+m_b)/\sqrt{2} $ & $m_t^2+m_b^2$ & $3$\\
\hline
$H^-$   & $\mu$ & $\bar{\nu}_\mu$ & $-m_\tau/\sqrt{2} $    & $\phantom{+}m_\tau/\sqrt{2} $ & $m_\tau^2$ & $1$\\
        & $s$   & $\bar{c}$   & $(m_t-m_b)/\sqrt{2}$ & $ (m_t+m_b)/\sqrt{2}$ & $m_t^2+m_b^2$ & $3$\\
\end{tabular}
\caption{\label{tab3} The fermionic decays of the Higgs particles in the MCPM
and the corresponding coupling constants $a$ and $b$ of Fig.~\ref{fig1}.}
\end{table*}
Here $\theta$ is the step function and 
\begin{equation}
\label{eq27}
w(x,y,z) = \big( x^2+y^2+z^2-2xy-2yz-2zx \big)^{1/2}
\end{equation}
is the usual kinematic function. Inserting in (\ref{eq26})
the values $a$ and $b$ from Tab. \ref{tab3} we get the results
for the individual decay rates as discussed below. For the fermion masses
we use the values from \cite{Amsler:2008zz}.

The rates for the decays of $\rho'$ to $t \bar{t}$ and $b \bar{b}$ are,
at tree level, as for the SM Higgs particle $\rho'_{SM}$. 
In the strict symmetry limit
of the MCPM, as we discuss it here, the first- and second-family
fermions are massless and $\rho'$ will not decay to them at tree level.
In reality this will, of course, be only an approximation.
In Nature we find very small but non-zero values for the ratios
of first- and second-family masses to the 
corresponding third family masses;
see (125) of \cite{Maniatis:2007de}. 
Thus, we should conclude that the Higgs particle $\rho'$
of the MCPM has the decays to second-- and first--family fermions
highly suppressed as is also the case for the SM Higgs boson $\rho'_{SM}$. 

For the Higgs particles $h'$, $h''$, $H^+$ and $H^-$ the dominant
fermionic decays are according to Tab. \ref{tab3}
\begin{equation}
\label{eq28}
\begin{split}
h'  \rightarrow& c \bar{c}\,,\\
h'' \rightarrow& c \bar{c}\,,\\
H^+ \rightarrow& c \bar{s}\,,\\
H^- \rightarrow& s \bar{c}\,.
\end{split}
\end{equation}
The numerical results for the decay widths
with the $s$ and $c$ quark masses
set to zero are
given in Tab. \ref{tab4}.
Note that
these partial decay widths are proportional to the respective Higgs-boson mass.
\begin{table}
\centering
\begin{tabular}{c|c}
decay & partial width $\Gamma$ [GeV]\\
\hline
$h'  \rightarrow c \bar{c}$ & $12.08$ $(m_{h'}/200 \text{ GeV})$\\
$h'' \rightarrow c \bar{c}$ & $12.08$ $(m_{h''}/200 \text{ GeV})$\\
$H^+ \rightarrow c \bar{s}$ & $12.09$ $(m_{H^\pm}/200 \text{ GeV})$\\
$H^- \rightarrow s \bar{c}$ & $12.09$ $(m_{H^\pm}/200 \text{ GeV})$\\
\end{tabular}
\caption{\label{tab4} 
The partial widths for the leading fermionic decays of $h'$, $h''$, $H^+$ and $H^-$.
The Higgs-boson masses have to be inserted in units of GeV.}
\end{table}

%%%%%%%%%%%%%%%%%%%%%%%%%%%%%%%%%%%%%%%%%%%%%%%%%%%%%%%%%%%%%%%%%%%%%%%%
% subsection III.2
%
%%%%%%%%%%%%%%%%%%%%%%%%%%%%%%%%%%%%%%%%%%%%%%%%%%%%%%%%%%%%%%%%%%%%%%%
\subsection[Decays into Higgs plus gauge boson]{Decays of a Higgs particle into another Higgs particle plus a gauge boson}
\label{secIII2}

\begin{figure}[t] 
\centering
\begin{tabular}{m{0.5\linewidth}m{0.4\linewidth}}
\includegraphics[width=.7\linewidth]{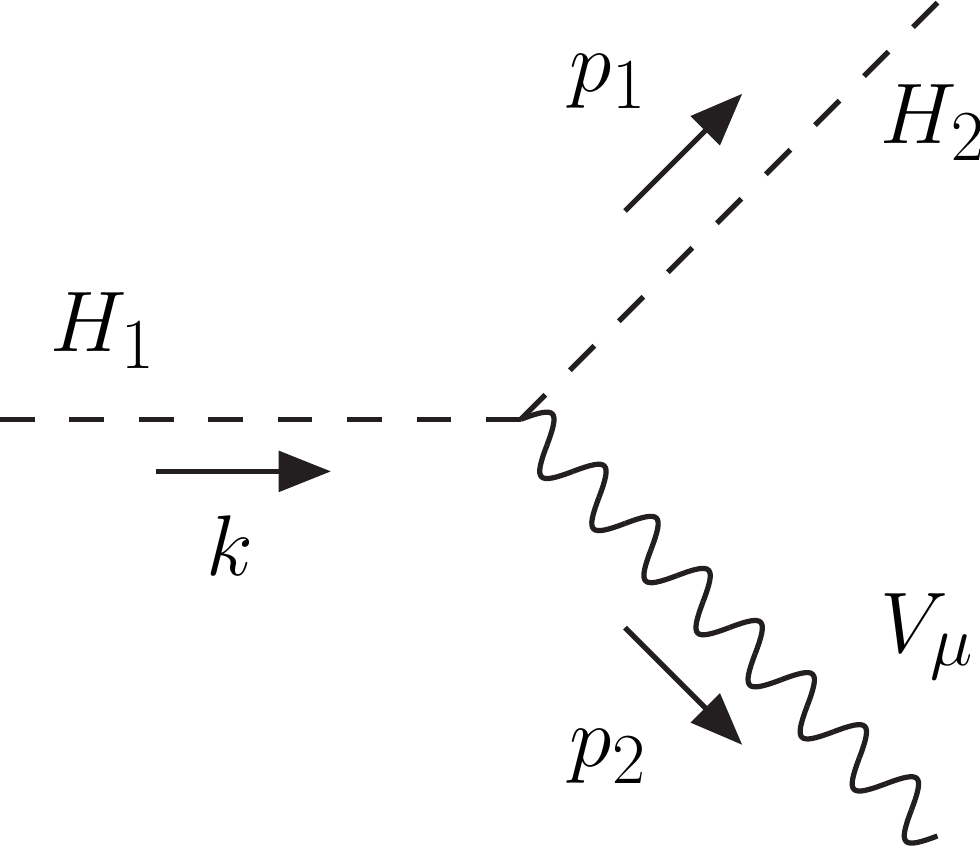} 
\hspace{0.1\linewidth}
&
{\large $i e \;C\; (k + p_1)_\mu$}
\\
\end{tabular}\\
\caption{\label{fig2}
The generic tree level diagram and vertex 
expression for the decay (\ref{eq29})}
\end{figure}

Here we discuss the decays
\begin{equation}
\label{eq29}
H_1(k) \rightarrow H_2(p_1) + V(p_2)\;,
\end{equation}
where $H_1$ and $H_2$ generically denote
Higgs particles and $V$ a gauge boson, 
$V= Z, W^\pm, \gamma$. In (\ref{eq29})
the momenta are indicated in brackets.
The decay (\ref{eq29}) can, of course,
only proceed if $m_{H_1} \ge m_{H_2} + m_V$.
The generic tree level diagram and the analytic
expression for the vertex
for the decay (\ref{eq29}) are shown in Fig.~\ref{fig2}.
In the MCPM $h'$ always has higher mass than $h''$; see (\ref{eq23}).
Also, no decays (\ref{eq29}) with $V=\gamma$ occur
at tree level. This leaves us with the decays shown in Tab.
\ref{tab5}, where we also list the corresponding
values for the coupling constant $C$ in the vertex diagram in Fig.~\ref{fig2}.
\begin{table}
\centering
\begin{tabular}{c||c|c|c}
$H_1$ & $H_2$ & $V$ & $C$\\
\hline
$h'$  & $h''$ & $Z$   & $-{i}/({2 s_W c_W})$\\
$h'$  & $H^+$ & $W^-$ & $-{1}/({2 s_W})$\\
$h'$  & $H^-$ & $W^+$ & $\phantom{+}{1}/({2 s_W})$\\
$h''$ & $H^+$ & $W^-$ & $-{i}/({2 s_W})$\\
$h''$ & $H^-$ & $W^+$ & $-{i}/({2 s_W})$\\
$H^+$ & $h'$ & $W^+$ & $-{1}/({2 s_W})$\\
$H^+$ & $h''$& $W^+$ & $\phantom{+}{i}/({2 s_W})$\\
$H^-$ & $h'$ & $W^-$ & $\phantom{+}{1}/({2 s_W})$\\
$H^-$ & $h''$& $W^-$ & $\phantom{+}{i}/({2 s_W})$\\
\end{tabular}
\caption{\label{tab5}
The decays (\ref{eq29}) occurring at tree
level in the MCPM 
if the masses satisfy $m_{H_1} > m_{H_2} +m_V$.
The last column gives
the corresponding coupling constant $C$ in Fig.~\ref{fig2}.
Here $s_W \equiv \sin \theta_W$ and $c_W\equiv \cos \theta_W$ 
denote the sine and the cosine of the weak
mixing angle, respectively}
\end{table}
The decay rate for the process (\ref{eq29}) is easily calculated:
\begin{multline}
\label{eq31}
\Gamma (H_1 \rightarrow H_2 + V) =\\
\frac{\alpha}{4} |C|^2\;
\theta(m_{H_1}-m_{H_2}-m_V)\;
m_{H_1} \left( \frac{m_{H_1}}{m_V} \right)^2\\
\left( 1- \frac{ (m_{H_2} + m_V)^2 }{m_{H_1}^2} \right)^{3/2}
\left( 1- \frac{ (m_{H_2} - m_V)^2 }{m_{H_1}^2} \right)^{3/2}\;.
\end{multline}
Here $\alpha=e^2/(4 \pi)$ is the fine structure constant. 
The coupling constants $C$ in~\eqref{eq31} are given in Tab.~\ref{tab5}. 

\begin{figure}[t] 
\centering
\includegraphics[width=\linewidth,clip]{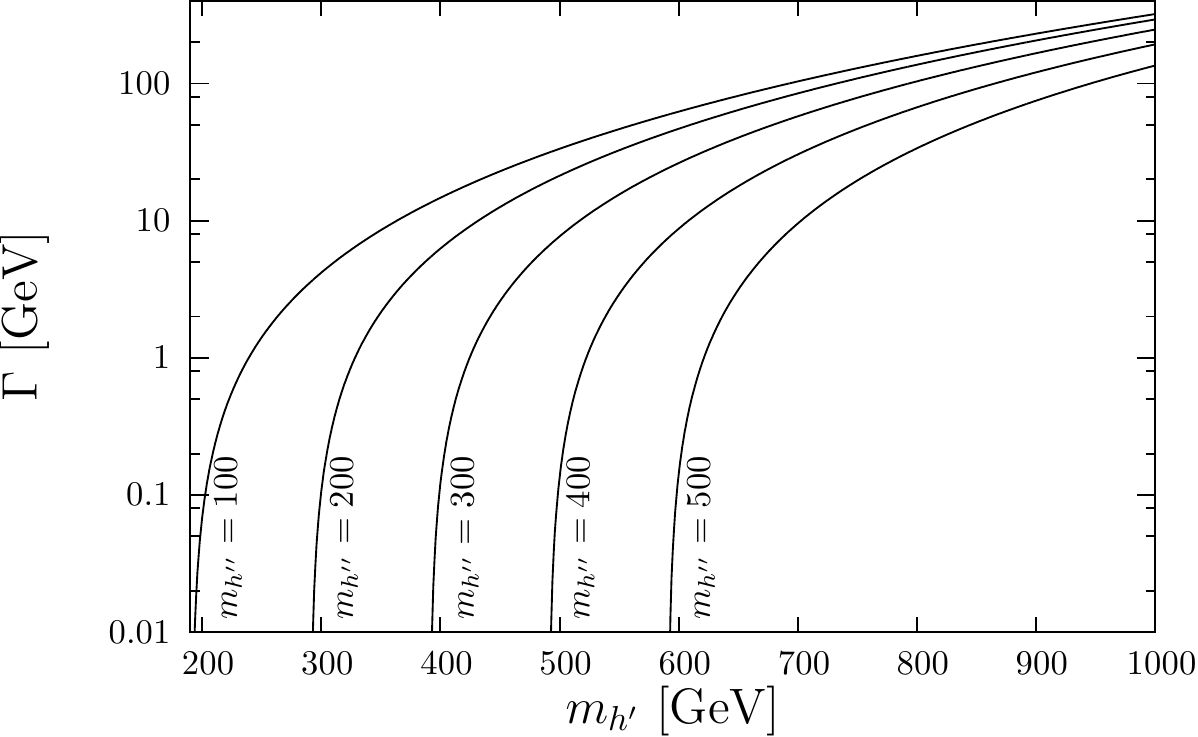}
\caption{\label{HHVdecay}
Partial width of the decay $h' \rightarrow h'' + Z$.
Shown is this width as function of the Higgs-boson
mass $m_{h'}$ for different fixed masses $m_{h''}$ from
$100$ to $500$~GeV in steps of $100$~GeV.}
\end{figure}

The partial width for the decay of the $h'$ boson into the $h''$ boson and an 
additional $Z$-boson is shown as function of the $h'$ mass
for fixed $h''$ masses in Fig.~\ref{HHVdecay}. We see that we 
get a width exceeding $10$~GeV only for rather 
large mass differences of the two
involved Higgs bosons. 
Considering for instance $m_{h''}= 100$~GeV we get from
Fig.~\ref{HHVdecay} $\Gamma > 10$~GeV only
for $m_{h'} > 364$~GeV.
For $m_{h''}= 300$~GeV, $\Gamma > 10$~GeV is reached
only for $m_{h'} > 517$~GeV. 
For the charged Higgs boson decays into a neutral Higgs boson
and a $W$ boson we get quite similar results. 

%

%%%%%%%%%%%%%%%%%%%%%%%%%%%%%%%%%%%%%%%%%%%%%%%%%%%%%%%%%%%%%%%%%%%%%%%%
% new subsection III.3a
%
%%%%%%%%%%%%%%%%%%%%%%%%%%%%%%%%%%%%%%%%%%%%%%%%%%%%%%%%%%%%%%%%%%%%%%%
\subsection{Decays of neutral Higgs bosons into a photon pair}
\label{secIII3gamma}

\begin{figure}[t] 
\centering
\begin{tabular}{m{0.5\linewidth}m{0.5\linewidth}}
\includegraphics[width=.7\linewidth]{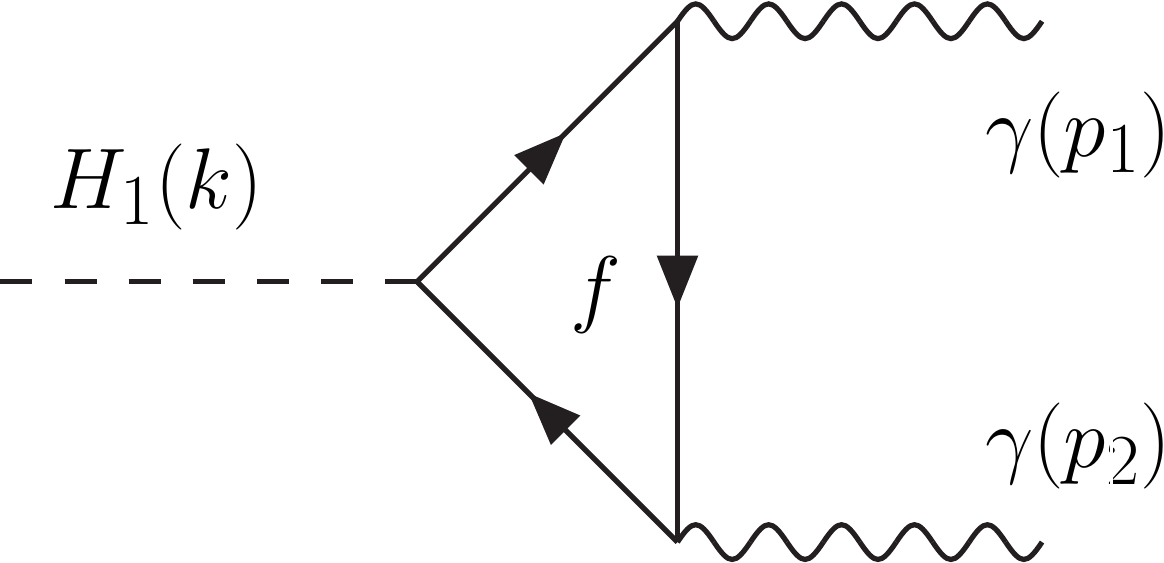} 
%\hspace{-0.1\linewidth}
&
{\large $+ \;\; \gamma(p_1) \leftrightarrow \gamma(p_2)$}
\\
\end{tabular}
\caption{\label{fHgamma}
Leading order diagrams for the decay $H_1\rightarrow \gamma \gamma$
(\ref{eqHgamma}) with a loop of a fermion~$f$.}
\end{figure}

Here we discuss the decays
\begin{equation}
\label{eqHgamma}
H_1(k) \rightarrow \gamma(p_1) + \gamma(p_2)
\end{equation}
in the MCPM where $H_1$ generically denotes a neutral
Higgs particle,
\begin{equation}
\label{eqnH}
H_1 = \rho', h', h'' \;.
\end{equation}
We have to consider in general contributions to the decay (\ref{eqHgamma})
via a fermion loop, a $W$-boson loop and a loop of a charged
Higgs boson. In Fig.~\ref{fHgamma} the Feynman diagram for
the contribution of a fermion loop is shown.

The couplings of the Higgs bosons~\eqref{eqnH} to the
fermions, the $W$-boson and the charged Higgs bosons
are given in the Feynman rules in
Appendix \ref{appA}.\\

For the calculation of the decay rate
for~\eqref{eqHgamma} we rely on the 
results of \cite{Gunion:1989we} which give
\begin{equation}
\label{eLHgamma}
\Gamma (H_1 \rightarrow \gamma + \gamma) =
\frac{\alpha^2}{256 \pi^3}
\frac{m_{H_1}^3}{v_0^2}
\left|
	\sum\limits_{i=f,W,H^\pm} I_{H_1}^i
\right|^2\,.
\end{equation}
The contributions of the various loops are as follows:
\begin{itemize}
\item fermion loops, 
\begin{equation}
\label{eq-IH1f}
I_{H_1}^f = 4 N_c^f\; e_f^2\; R_f^{H_1}/m_{H_1}^2\; 
F^{H_1}_{\frac{1}{2}}(\frac{4 m_f^2}{m_{H_1}^2})
\end{equation}
with $e_f$ the charge of the fermion in units of the positron charge, $N_c^f$ the
color factor and
\begin{equation}
\label{eqRq}
R_f^{H_1}=
\begin{cases}
\phantom{+} m_f^2, &\text{ for } H_1= \rho', f=t,b, \tau \\
- m_t m_c, &\text{ for } H_1= h', f=c\\
- m_b m_s, &\text{ for } H_1= h', f=s\\
- m_\tau m_\mu, &\text{ for } H_1= h', f=\mu\\
- m_t m_c, &\text{ for } H_1= h'', f=c\\
\phantom{+} m_b m_s, &\text{ for } H_1= h'', f=s\\
\phantom{+} m_\tau m_\mu, &\text{ for } H_1= h'', f=\mu\\
\phantom{+} 0, &\text{ otherwise }\,.
\end{cases}
\end{equation}
Furthermore we set
\begin{equation}
\label{eqf12}
F^{H_1}_{\frac{1}{2}}(z)= 
\begin{cases}
-2  \big[ 1+ (1-z) f(z) \big], \quad & \text{for } H_1=\rho', h'\\
-2  f(z), \quad & \text{for } H_1=h''\;.
\end{cases}
\end{equation}
\item $W$-boson loop, 
\begin{equation}
\label{eq-Wl}
I_{h'}^W=0,\quad I_{h''}^W=0,\quad
I_{\rho'}^W = F_1(\frac{4 m_W^2}{m_{\rho'}^2})
\end{equation}
with
\begin{equation}
F_1(z) = 2 + 3  z + 3 z (2-z) f(z)\;.
\end{equation}
\item $H^\pm$-boson loop, 
\begin{equation}
\label{eq-Hpmloop}
I_{h'}^{H^\pm}=0,\quad I_{h''}^{H^\pm}=0,\quad
I_{\rho'}^{H^\pm} = \frac{m_{\rho'}^2+2 m_{H^\pm}^2}{2 m_{H^\pm}^2} F_0(\frac{4 m_{H^\pm}^2}{m_{\rho'}^2})
\end{equation}
with
\begin{equation}
\label{eq-F0}
F_0(z) = z \big[ 1 - z  f(z) \big]\;.
\end{equation}
\end{itemize}
Finally, $f(z)$ is defined as
\begin{equation}
f(z) = 
\begin{cases}
-\frac{1}{4}
\bigg[ \ln \bigg( \frac{1+ \sqrt{1-z}}{1- \sqrt{1-z}}\bigg) - i \pi \bigg]^2,
\quad & \text{for } 0<z<1 \\
\big[ \arcsin ( \sqrt{1/z} ) \big]^2,
\quad & \text{for } z \ge 1 \;.
\end{cases} 
\end{equation}

\begin{figure}[t] 
\centering
\includegraphics[width=\linewidth,clip]{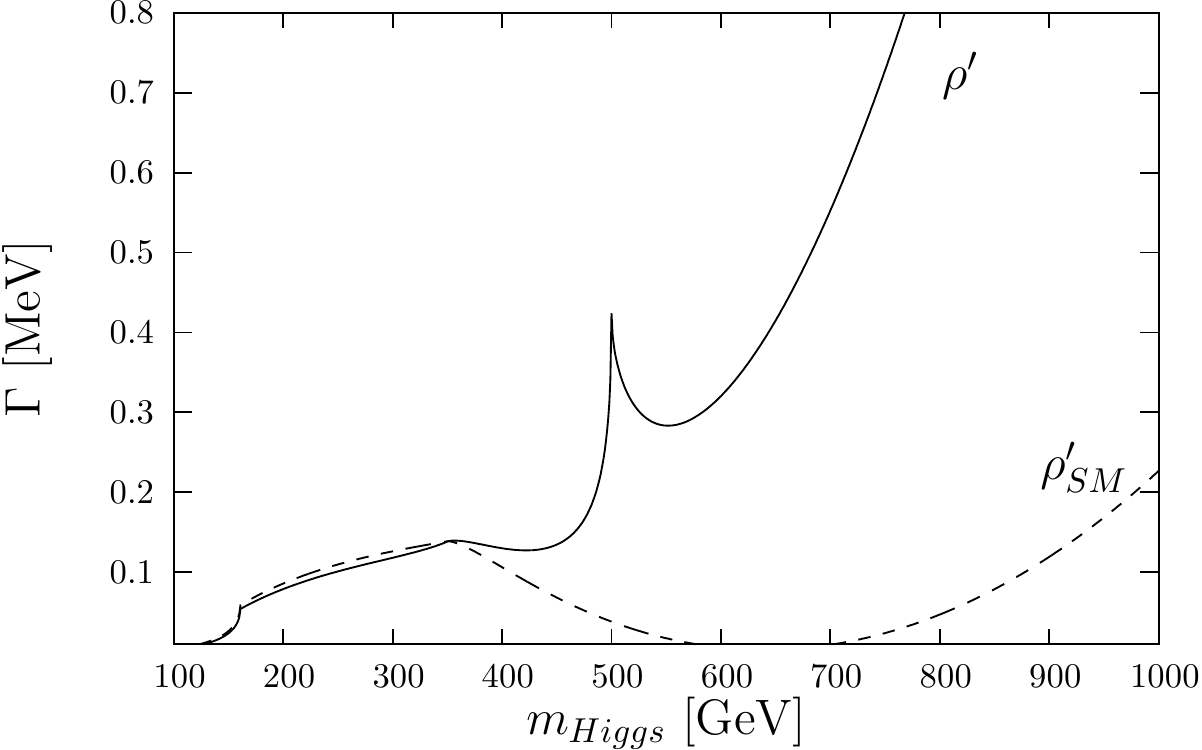}
\caption{\label{Hggdecay} Partial decay widths of the
neutral Higgs boson $\rho'$ and of the
SM Higgs boson $\rho'_{\text SM}$ into a pair of photons.
The charged Higgs-boson mass is supposed to be $m_{H^\pm}= 250$~GeV.}
\end{figure}

For the decays of the neutral Higgs bosons into a photon pair we find
only small widths from these results.
The partial decay width of the $\rho'$ boson is compared to
the corresponding width of the $\rho'_{SM}$ in Fig.~\ref{Hggdecay}.
We get significant deviations of the $2 \gamma$ decay widths of the 
$\rho'$ boson and
the SM boson $\rho'_{SM}$ only if  
$m_{\rho'}$ is near to or higher than
twice the charged Higgs-boson mass which we
set to $m_{H^\pm} = 250$~GeV in this plot. 
Of course, the
peak at twice the charged Higgs-boson mass is an artifact due
to our neglect of the finite width of $H^\pm$ in the calculation.
The peak will become a broader structure if
the non-vanishing $H^\pm$ width is taken into account.
Let us note that even for large charged Higgs-boson masses
the corresponding loop contribution does not decouple.
This comes about as follows. Consider the diagram of Fig.~\ref{fHgamma}
with a $H^\pm$ loop instead of the fermion loop $f$. The $\rho' H^+H^-$
coupling contains a factor $m_{H^\pm}^2$; see \eqref{eqA2} 
in appendix \ref{appA}.
The loop integration gives for large $m_{H^\pm}^2$,
using simple power counting arguments, a factor $1/m_{H^\pm}^2$.
The net result is a finite contribution to the amplitude $\rho'\rightarrow \gamma \gamma$
even for large $m_{H^\pm}^2$. This is, of course, borne out by the
explicit calculation in \eqref{eq-Hpmloop} from which we find
\begin{equation}
I_{\rho'}^{H^\pm} \rightarrow -\frac{1}{3}
\end{equation}
for $m_{H^\pm} \rightarrow \infty$ keeping $m_{\rho'}$ fixed.\\

For masses of the $\rho'_{SM}$ boson
of 120 to 150~GeV the decay channel
$\rho'_{SM} \rightarrow \gamma \gamma$ is an
important discovery mode at the LHC. We find here
that the $2 \gamma$ width of $\rho'_{SM}$ 
and of the $\rho'$ in the MCPM are quite similar
for this mass range if 
$m_{H^\pm} > 200$~GeV. As we shall
show below in Sect.~\ref{IV2} also
the production cross sections for $\rho'$
and $\rho'_{SM}$ are practically equal.
Thus, in the above mass range, the $2 \gamma$ channel is
as good a discovery channel for $\rho'$ as it is for $\rho'_{SM}$.\\

We turn now to the $2 \gamma$ decays of $h'$ and $h''$.
We see from \eqref{eq-IH1f},\eqref{eq-Wl} and \eqref{eq-Hpmloop} that here
only the fermion loops contribute. This comes about
since there are no couplings linear in $h'$ or $h''$ to
a $W^+ W^-$ and a $H^+ H^-$ pair in the MCPM; see 
\eqref{eqA2} of appendix \ref{appA}.
The only fermion flavors which contribute at one loop level to the
$ 2 \gamma$ decays of 
$h'$ and $h''$ are the $c$ and $s$ quarks
and the muon $\mu$. 
In the strict symmetry limit of the MCPM these fermions are massless.
Of course, in reality they get masses. Thus we have kept these
masses in the loop calculation.
The structure of the results can be seen from~\eqref{eLHgamma}-\eqref{eqf12}.
Let us consider as an example the $c$-quark-loop contribution
to $h' \rightarrow \gamma \gamma$. The factor
$I_{h'}^c$ \eqref{eq-IH1f} contains $R_c^{h'}= -m_t m_c$ where
$m_t$ originates from the $h' c \bar{c}$ vertex (see the Feynman rules in the appendix),
whereas $m_c$ comes from the loop integration. The term
$F^{h'}_{\frac{1}{2}}(4 m_c^2/m_{h'}^2)$ is
proportional to $\ln^2(m_{h'}/m_c)$ for $m_c\rightarrow 0$. Thus, $I_{h'}^c$ vanishes
for $m_c\rightarrow 0$. For the muon and $s$-quark loops the discussion is analogous.
In the strict symmetry limit of the MCPM where $m_c=m_s=m_\mu=0$ we have,
therefore, 
$\Gamma (h' \rightarrow \gamma \gamma) =\Gamma (h'' \rightarrow \gamma \gamma) = 0$.
In order to get a reasonable estimate for these rates we keep
the finite fermion masses in the loop calculation. This estimate gives tiny partial rates.
For the $h'$ and $h''$ decays into
a photon pair we find partial widths rising to about 3.5~keV for
Higgs-boson masses increasing from zero up to 35~GeV.
For Higgs-boson masses higher than 35~GeV the partial widths decrease
monotonically with increasing masses. Thus, these partial widths
are never larger than 3.5 keV which is very small compared to the 
decay widths of the main fermionic modes of Tab.~\ref{tab4}.

%%%%%%%%%%%%%%%%%%%%%%%%%%%%%%%%%%%%%%%%%%%%%%%%%%%%%%%%%%%%%%%%%%%%%%%%
% subsection III.3
%
%%%%%%%%%%%%%%%%%%%%%%%%%%%%%%%%%%%%%%%%%%%%%%%%%%%%%%%%%%%%%%%%%%%%%%%
\subsection{Decays of neutral Higgs bosons into two gluons}
\label{secIII3}

\begin{figure}[t] 
\centering
\begin{tabular}{m{0.5\linewidth}m{0.5\linewidth}}
\includegraphics[width=.7\linewidth]{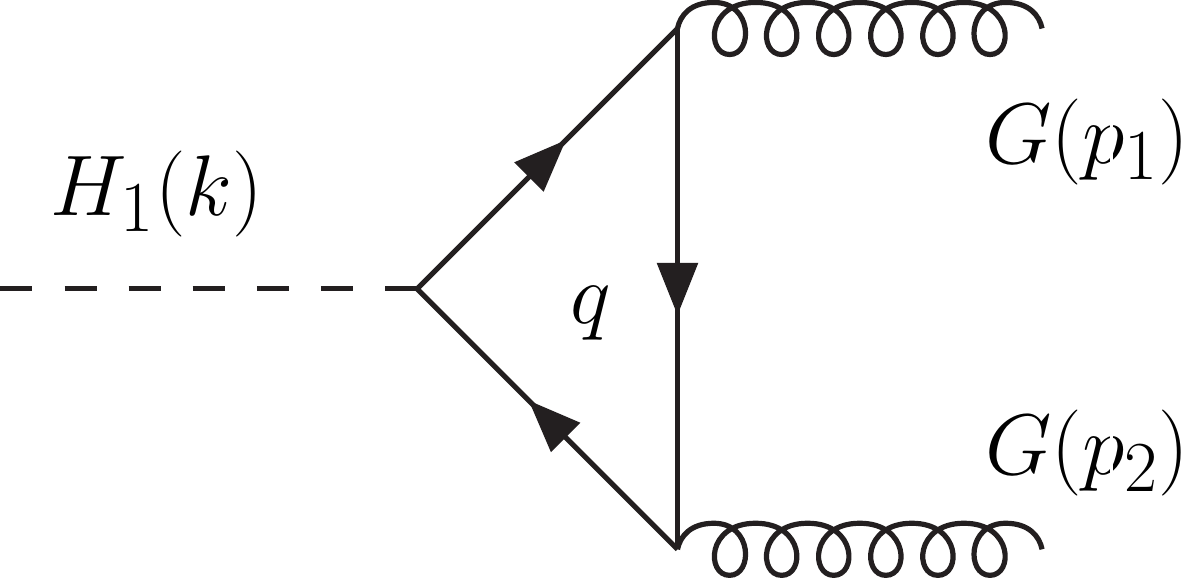} 
%\hspace{-0.1\linewidth}
&
{\large $+ \;\; G(p_1) \leftrightarrow G(p_2)$}
\\
\end{tabular}
\caption{\label{fig3}
Leading order diagrams for the decay $H_1\rightarrow GG$
(\ref{eq35}) with one loop of a quark $q$.}
\end{figure}

Here we discuss the decays
\begin{equation}
\label{eq35}
H_1(k) \rightarrow G(p_1) + G(p_2)
\end{equation}
in the MCPM where $H_1$ generically denotes a neutral
Higgs particle \eqref{eqnH}. 
The leading contributions to the decay (\ref{eq35})
proceed via quark loops as shown in Fig.~\ref{fig3}.

The calculation of the diagrams of Fig.~\ref{fig3} is quite analogous 
to that for the two-photon decay with
an internal quark loop; see Fig.~\ref{fHgamma}. 
Of course, in the gluon pair decay
there are no contributions
of a $W$-boson and a $H^\pm$ in the loop. Replacing 
$\alpha$ by the strong coupling parameter $\alpha_s$ and changing the color factor
appropriately we get
\begin{equation}
\label{eLHgluon}
\Gamma (H_1 \rightarrow G + G) =
\frac{\alpha_s^2}{128 \pi^3}
\frac{m_{H_1}^3}{v_0^2}
\left|
	\sum\limits_{q=c,s,t,b} \tilde{I}_{H_1}^q
\right|^2
\end{equation}
with
\begin{equation}
\label{eqIgluon}
\tilde{I}_{H_1}^q = (4 R_q^{H_1}/m_{H_1}^2)\; F^{H_1}_{\frac{1}{2}}(4 m_q^2/m_{H_1}^2)
\end{equation}
and the factors $R^{H_1}_q$ from~\eqref{eqRq} for the
different contributions of the quark flavors and
the function $F^{H_1}_{\frac{1}{2}}$ from \eqref{eqf12}.
Explicitly \eqref{eLHgluon} reads for
the neutral Higgs bosons $\rho'$, $h'$ and $h''$: 
\begin{equation}
\begin{split}
\label{eq37}
&\Gamma (\rho' \rightarrow G + G) =\\
&\frac{\alpha_s^2}{128 \pi^3}
\frac{m_{\rho'}^3}{v_0^2}
\left|
	4 \frac{m_t^2}{m_{\rho'}^2}  F^{\rho'}_{\frac{1}{2}}\bigg(\frac{4 m_t^2}{m_{\rho'}^2}\bigg)
	+ 4 \frac{m_b^2}{m_{\rho'}^2}  F^{\rho'}_{\frac{1}{2}}\bigg(\frac{4 m_b^2}{m_{\rho'}^2} \bigg)
\right|^2 \,,
\end{split}
\end{equation}
\begin{equation}
\begin{split}
\label{eq38}
&\Gamma (h' \rightarrow G + G) =\\
&\frac{\alpha_s^2}{128 \pi^3}
\frac{m_{h'}^3}{v_0^2}
\left|
	4 \frac{m_t m_c}{m_{h'}^2}  F^{h'}_{\frac{1}{2}}\bigg(\frac{4 m_c^2}{m_{h'}^2}\bigg)
	+4 \frac{m_b m_s}{m_{h'}^2}  F^{h'}_{\frac{1}{2}}\bigg(\frac{4 m_s^2}{m_{h'}^2}\bigg)
\right|^2 
\end{split}
\end{equation}
and
\begin{equation}
\begin{split}
\label{eq39}
&\Gamma (h'' \rightarrow G + G) =\\
&\frac{\alpha_s^2}{128 \pi^3}
\frac{m_{h''}^3}{v_0^2}
\left|
	4 \frac{m_t m_c}{m_{h''}^2}  F^{h''}_{\frac{1}{2}}\bigg(\frac{4 m_c^2}{m_{h''}^2}\bigg)
	-4 \frac{m_b m_s}{m_{h''}^2}  F^{h''}_{\frac{1}{2}}\bigg(\frac{4 m_s^2}{m_{h''}^2}\bigg)
\right|^2.
\end{split}
\end{equation}
For the numerics we take the strong coupling at the $Z$-mass scale, 
$\alpha_s=0.12$ and $m_t= 171$~GeV.

Now we discuss the result \eqref{eLHgluon}-\eqref{eq39}.
Let us first consider the decay rate for 
$\rho' \rightarrow G + G$.
The one-loop
contributions from the $t$ and $b$ quarks are identical
to the corresponding SM expressions. In the strict symmetry limit
of the MCPM the other quarks,
$c$, $s$, $u$ and $d$ are massless and do not contribute
to $\rho' \rightarrow GG$ at one loop level. In reality
we thus expect that their contribution is very small. The same
is true in the SM where the $c$, $s$, $u$ and $d$ quarks 
together give only a $0.005\%$ contribution to
the decay width for $\rho'_{SM} \rightarrow GG$. Thus we find that in the MCPM
the decay rate for $\rho' \rightarrow GG$ is practically as 
in the SM for $\rho'_{SM}$.

\begin{figure}[t] 
\centering
\includegraphics[width=\linewidth,clip]{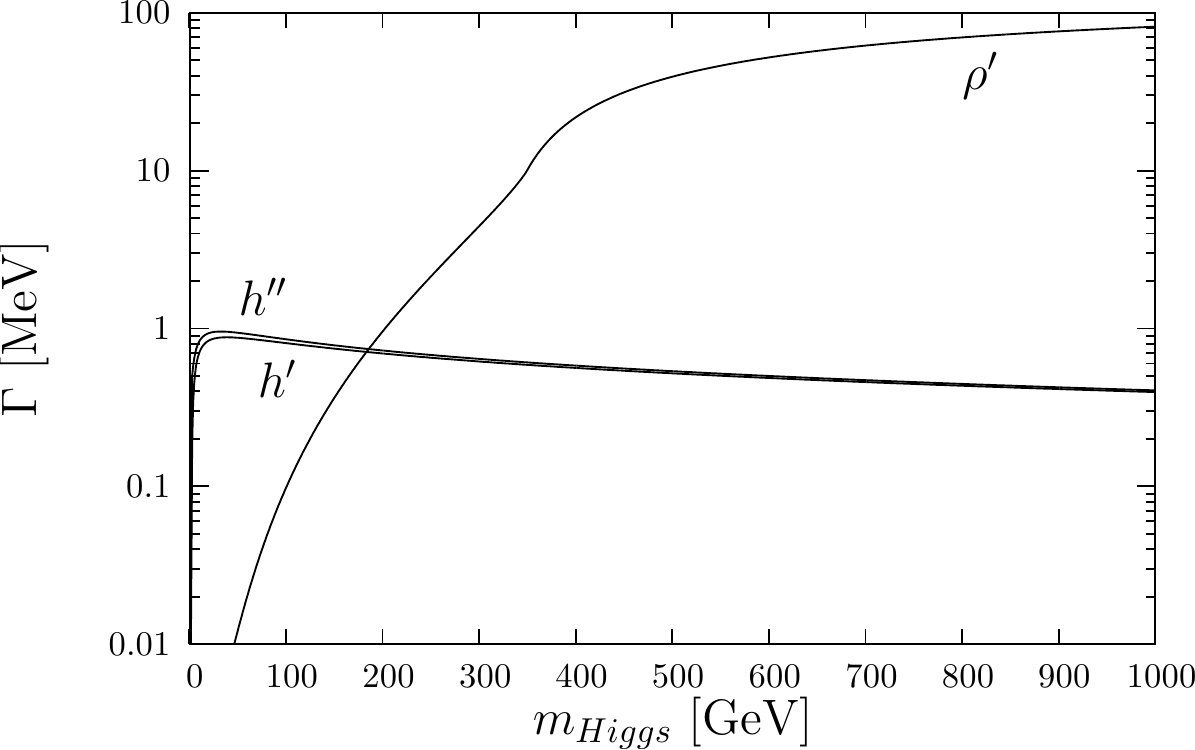}
\caption{\label{HGGdecay} Partial decay widths of the
neutral Higgs bosons into a pair of gluons.}
\end{figure}

Turning now to the decays $h' \rightarrow GG$ and
$h'' \rightarrow GG$ we must clearly say that in the strict
symmetry limit of the MCPM where $m_c=m_s=0$
we have $\Gamma(h' \rightarrow GG)=\Gamma(h'' \rightarrow GG)=0$; 
see \eqref{eq38} and \eqref{eq39}.
But we can argue that in reality $m_c$ and $m_s$ are unequal
to zero. Then, the Higgs particles $h'$ and $h''$ with the
couplings to $c$ and $s$ quarks given in Appendix \ref{appA}
will indeed decay into two gluons.
The dominant contributions 
come from the $c$ quark loops since
the couplings of $h'$ and $h''$ to $c$ quarks are
proportional to the large $t$-quark mass. 
But even with this enhancement factor we find 
only partial widths of the
order of MeV for the decays $h' \rightarrow GG$ and
$h'' \rightarrow GG$, respectively; see 
Fig.~\ref{HGGdecay}.
Comparing with the results for the dominant fermionic
decay modes of $h'$ and $h''$ as shown in Tab. \ref{tab4}
we find that the branching ratios for
$h' \rightarrow G G$ and
$h'' \rightarrow G G$  are predicted to be less
than about  $10^{-4}$. 
Nevertheless, the results for the gluonic decays (\ref{eq35}) will be
needed for the discussion of the Higgs-boson production processes 
in the following section.

We summarize our findings for the Higgs-boson decays. 

Firstly, we have results valid in the strict symmetry limit.
We find that the $\rho'$ decays 
are in essence as for the SM Higgs boson $\rho'_{SM}$. 
Only if $m_{\rho'}$ comes near to or is larger than $ 2 m_{H^\pm}$
we do find large deviations between $\Gamma(\rho' \rightarrow \gamma \gamma)$
and $\Gamma(\rho'_{SM} \rightarrow \gamma \gamma)$.
If the Higgs particles $h'$, $h''$ and $H^\pm$ have masses
below about $400$~GeV their main decays are the fermionic ones 
as given in \eqref{eq28} and Tab.~\ref{tab4}.
These rates can
be taken as good estimates for the total decay rates of
$h'$, $h''$ and $H^\pm$, respectively. From (\ref{eq26}) and 
Tab. \ref{tab3}
we can estimate the branching ratios for the decays into leptons of
the second family as
\begin{equation}
\label{eq-branch}
\begin{split}
&\frac{\Gamma(h' \rightarrow \mu^- \mu^+)}{\Gamma(h' \rightarrow \text{all})}
\approx 
\frac{\Gamma(h'' \rightarrow \mu^- \mu^+)}{\Gamma(h'' \rightarrow \text{all})} \approx\\
 &
\frac{\Gamma(H^+ \rightarrow \mu^+ \nu_\mu)}{\Gamma(H^+ \rightarrow \text{all})}
\approx 
\frac{\Gamma(H^- \rightarrow \mu^- \bar{\nu}_\mu)}{\Gamma(H^- \rightarrow \text{all})} \approx\\
 & \frac{m_\tau^2}{3(m_t^2+m_b^2) + m_\tau^2} 
\approx  3 \times 10^{-5} \,.
\end{split}
\end{equation}
In the symmetry limit the Higgs particles $h'$, $h''$ and $H^\pm$ do not couple
to the fermions of the first and third families. Thus, the branching ratios
for the decays of the Higgs-bosons $h'$, $h''$ and $H^\pm$ to leptons of the first and third 
families are predicted
to be very small in the MCPM. 
Note that this predicted large suppression
of the decay modes involving $\tau$ and $\nu_\tau$ leptons relative
to the modes involving $\mu$ and $\nu_\mu$ is a feature of the MCPM which
distinguishes it from more conventional THDMs.

\begin{figure}[t] 
\centering
\includegraphics[width=\linewidth,clip]{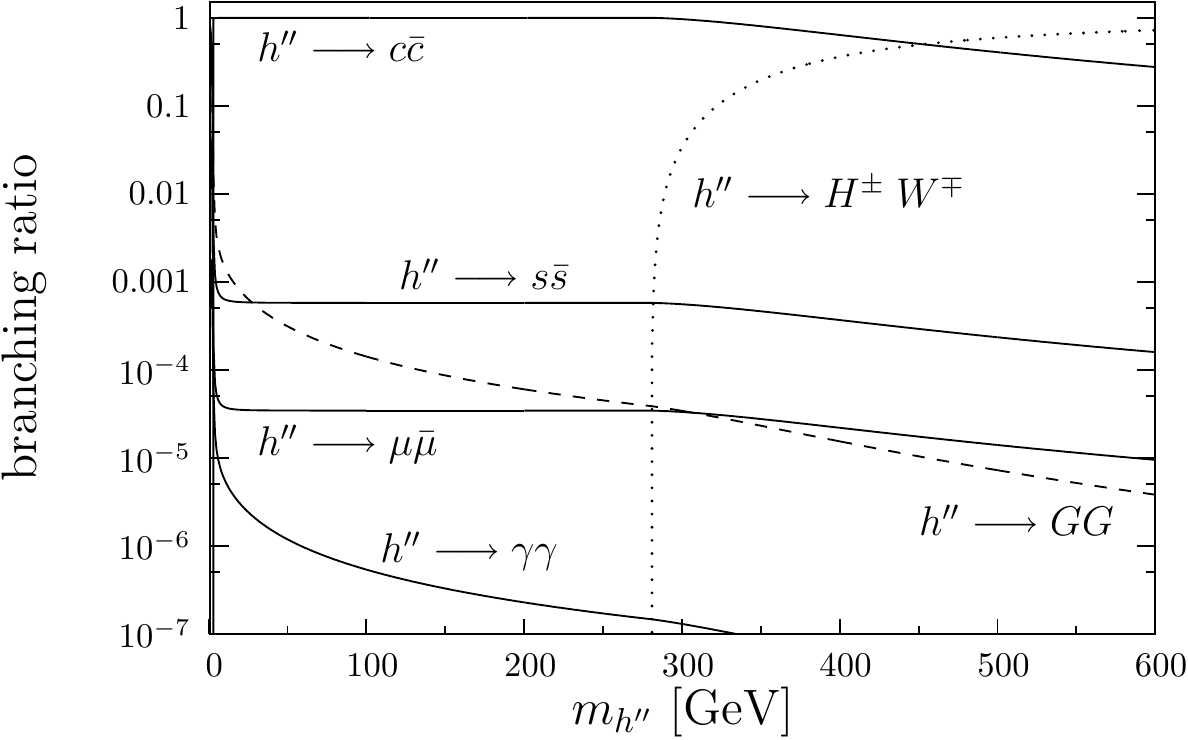}
\caption{\label{Hbranch} Branching fractions
for the $h''$ Higgs-boson decays into different available decay channels
as functions of $m_{h''}$. It is supposed that $m_{H^\pm}=200$~GeV.
The curve for $h'' \rightarrow H^\pm W^\mp$ corresponds to the sum
of these two channels.
}
\end{figure}

\begin{figure}[t] 
\centering
\includegraphics[width=\linewidth,clip]{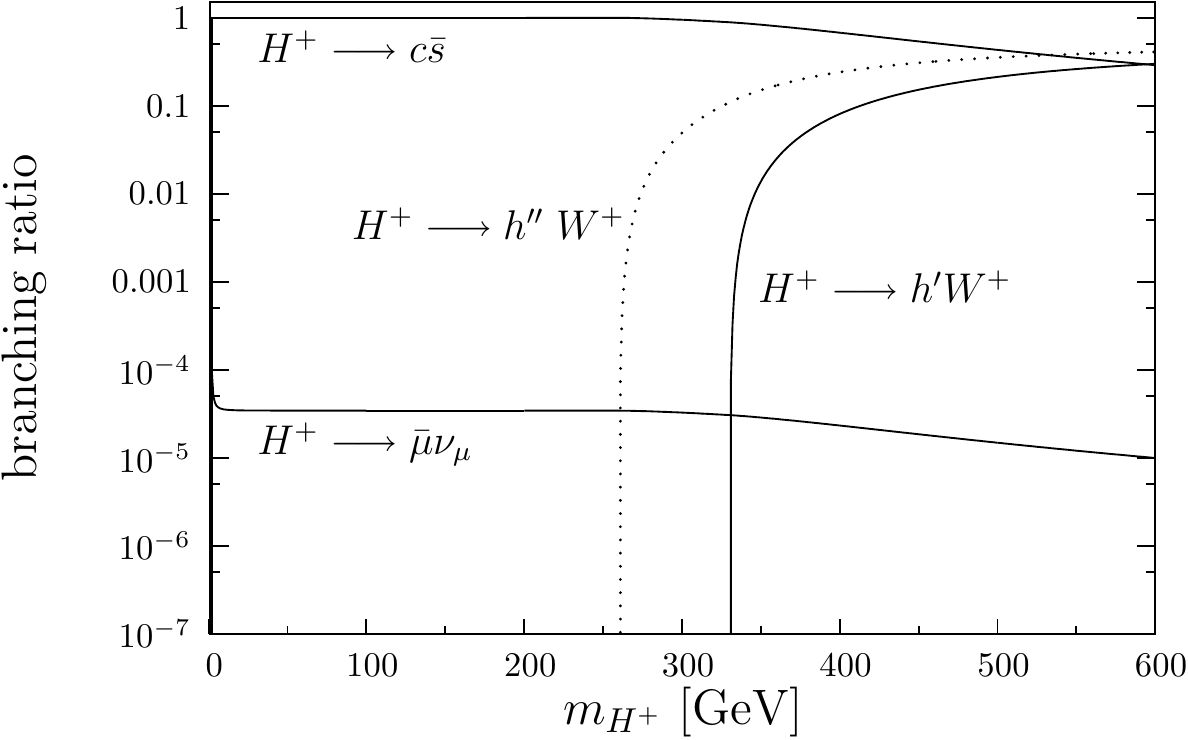}
\caption{\label{Hpmbranch} Branching fractions
for the $H^+$ Higgs-boson decay channels as function of $m_{H^+}$.
It is supposed that $m_{h'}=250$~GeV and 
$m_{h''}=180$~GeV.}
\end{figure}

Secondly, we have estimates going beyond the strict symmetry limit,
where the masses of the second- and first-family fermions
are zero. In the strict limit the decay rates
$\Gamma(h' \rightarrow \gamma \gamma)=\Gamma(h'' \rightarrow \gamma \gamma)
=\Gamma(h' \rightarrow GG)=\Gamma(h'' \rightarrow GG)=0$.
Of course, in reality these decay rates will be non-zero. We
have given {\em estimates} for these decay rates using the physical
values for the masses of the second- and first-family fermions in
the corresponding loop calculations. These estimates
give very small values for the above decay rates which,
therefore, do not change the overall picture significantly.
As an example we show in
Fig.~\ref{Hbranch} the branching ratios for the $h''$ 
Higgs-boson decays 
for the channels $c\bar{c}$, $s\bar{s}$, $\mu \bar{\mu}$, $H^\pm W^\mp$, $GG$
and $\gamma \gamma$.
It is supposed that the
charged Higgs bosons $H^\pm$ have a mass 
of $200$~GeV.  
As another example we show in Fig.~\ref{Hpmbranch}
the branching ratios for the decays of the $H^+$ boson
as function of its mass $m_{H^+}$ supposing
$m_{h'}=250$~GeV and 
$m_{h''}=180$~GeV.

%%%%%%%%%%%%%%%%%%%%%%%%%%%%%%%%%%%%%%%%%%%%%%%%%%%%%%%%%%%%%%%%%%%%%%%%
% Higgs production
%
%%%%%%%%%%%%%%%%%%%%%%%%%%%%%%%%%%%%%%%%%%%%%%%%%%%%%%%%%%%%%%%%%%%%%%%
\section{Higgs-boson production}
\label{secIV}

In this section we shall discuss the production of Higgs particles
in proton--proton collisions at LHC energies. We write
generically
\begin{equation}
\label{eq50}
p (p_1) + p(p_2) \rightarrow H_1(k) + X\;,
\end{equation}
where $H_1$ denotes one of the Higgs particles of the MCPM;
$H_1=\rho', h', h'', H^\pm$. There are, of course, many
contributions to (\ref{eq50}). For a discussion of the contributions
to $\rho'_{SM}$ production in the framework of the SM see
for instance \cite{Buttar:2006zd}.

We shall focus here on two different Higgs-boson production mechanisms in
the MCPM, the quark--antiquark fusion and the gluon--gluon fusion.
As we shall see, we get in both cases results which are quite
distinct from those obtained in more conventional THDMs; see for instance 
\cite{Kim:2006tx}.

%%%%%%%%%%%%%%%%%%%%%%%%%%%%%%%%%%%%%%%%%%%%%%%%%%%%%%%%%%%%%%%%%%%%%%%%
% subsection IV.1
%
%%%%%%%%%%%%%%%%%%%%%%%%%%%%%%%%%%%%%%%%%%%%%%%%%%%%%%%%%%%%%%%%%%%%%%%
\subsection{Higgs-boson production by quark-antiquark fusion}
\label{IV1}

\begin{figure}[t] 
\centering
\includegraphics[width=0.7\linewidth,clip]{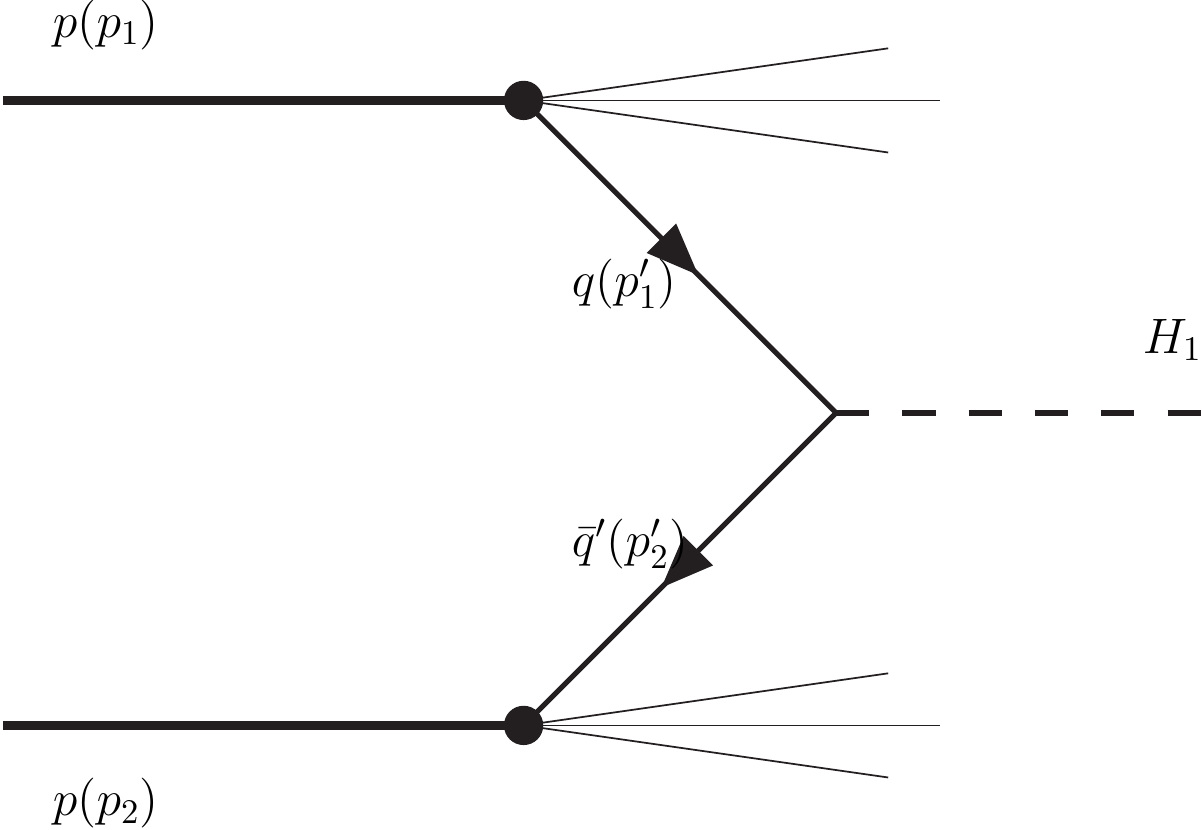}
\caption{\label{fig4}
The generic diagram for the production of 
a Higgs particle $H_1$ via quark--antiquark
fusion, $q \bar{q}' \rightarrow H_1$,
in proton--proton collisions.}
\end{figure}

\begin{figure}[h] 
\centering
\begin{tabular}{m{0.5\linewidth}m{0.4\linewidth}}
\includegraphics[width=.7\linewidth]{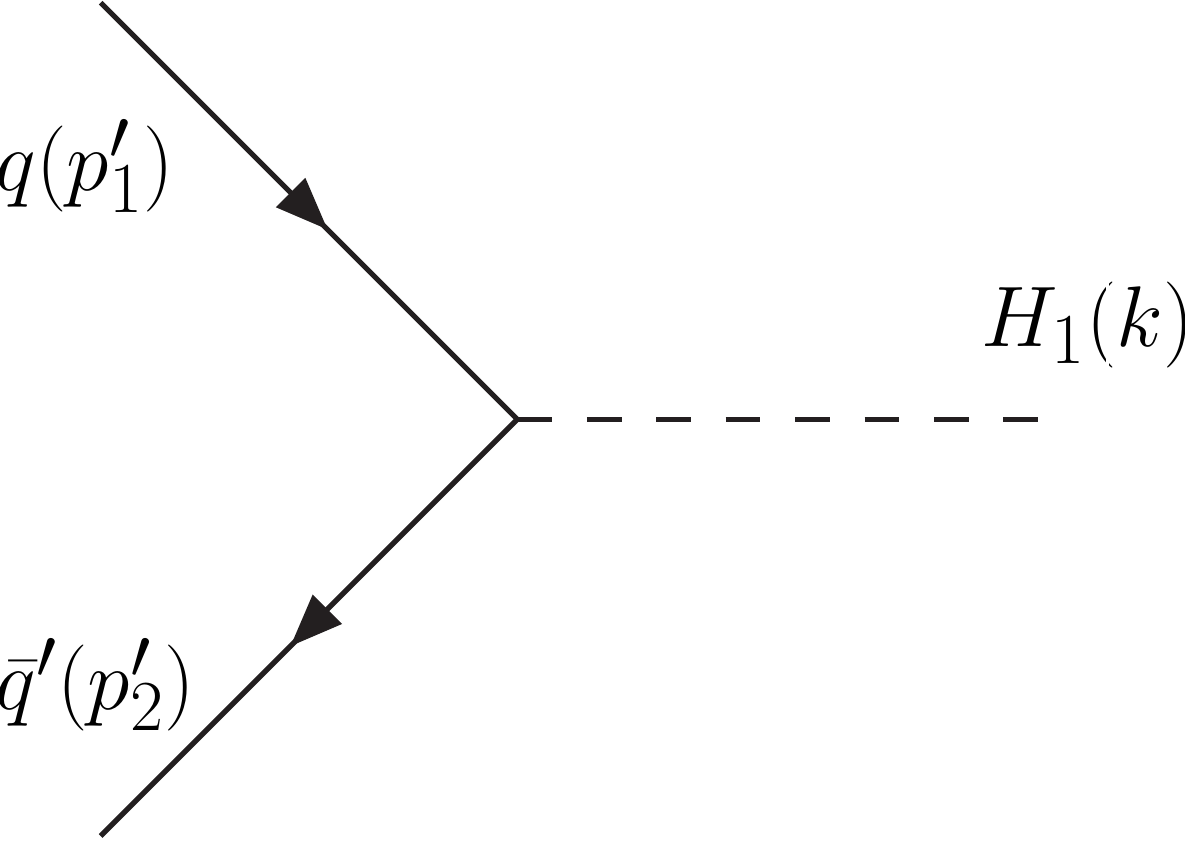} 
\hspace{0.1\linewidth}
&
{\large $-i \frac{1}{v_0} (a + b\; \gamma_5)$}
\\
\end{tabular}\\
\caption{\label{fig5}
The generic diagram for the fusion process
$q \bar{q}' \rightarrow H_1$ and
the corresponding analytic expression for the vertex.}
\end{figure}

Here we investigate the contribution to (\ref{eq50}) from
the quark-antiquark fusion, that is, the 
Drell-Yan type process. The generic diagram is
shown in Fig.~\ref{fig4}. The fusion processes
which can occur in the MCPM are listed in Tab. \ref{tab6}
together with the coupling constants $a$ and $b$ in the diagram
shown for the generic process in Fig.~\ref{fig5},
\begin{equation}
\label{eq51}
q(p_1') + \bar{q}'(p_2') \rightarrow H_1(k) \;.
\end{equation}
\begin{table}
\begin{tabular}{c||cc|cc|c}
$H_1$ & $q$ & $\bar{q}'$ & $a$ & $b$ & $|a|^2+|b|^2$\\
\hline
$\rho'$ & $t$ & $\bar{t}$ & $\phantom{+}m_t$ & 0 & $m_t^2$\\
	    & $b$ & $\bar{b}$ & $\phantom{+}m_b$ & 0 & $m_b^2$\\
\hline
$h'$    & $c$ & $\bar{c}$ & $-m_t$ & 0 & $m_t^2$\\
	    & $s$ & $\bar{s}$ & $-m_b$ & 0 & $m_b^2$\\
\hline
$h''$   & $c$ & $\bar{c}$ & $0$ & $\phantom{+}i m_t$& $m_t^2$\\
	    & $s$ & $\bar{s}$ & $0$ & $-i m_b$ & $m_b^2$\\
\hline
$H^+$   & $c$ & $\bar{s}$ & $\frac{1}{\sqrt{2}}(m_t-m_b)$ & $\phantom{+}\frac{1}{\sqrt{2}}(m_t+m_b)$ & $m_t^2+m_b^2$\\
$H^-$   & $s$ & $\bar{c}$ & $\frac{1}{\sqrt{2}}(m_t-m_b)$ & $-\frac{1}{\sqrt{2}}(m_t+m_b)$ & $m_t^2+m_b^2$\\
\end{tabular}
\caption{\label{tab6} 
The quark--antiquark fusion processes contributing
to the Higgs-boson production (\ref{eq50}) in the
MCPM and the corresponding coupling constants 
in Fig.~\ref{fig5}.}
\end{table}
For the $\rho'$ we have a large coupling to the $t$ quark.
But even at LHC energies there are not many $t$ and $\bar{t}$
quarks in the proton. Thus $\rho'$ production via
quark--antiquark fusion is unimportant in the MCPM.
This conclusion is exactly as in the SM for $\rho'_{SM}$; 
see for instance \cite{Buttar:2006zd}.

\begin{figure}[t] 
\centering
\includegraphics[width=\linewidth,clip]{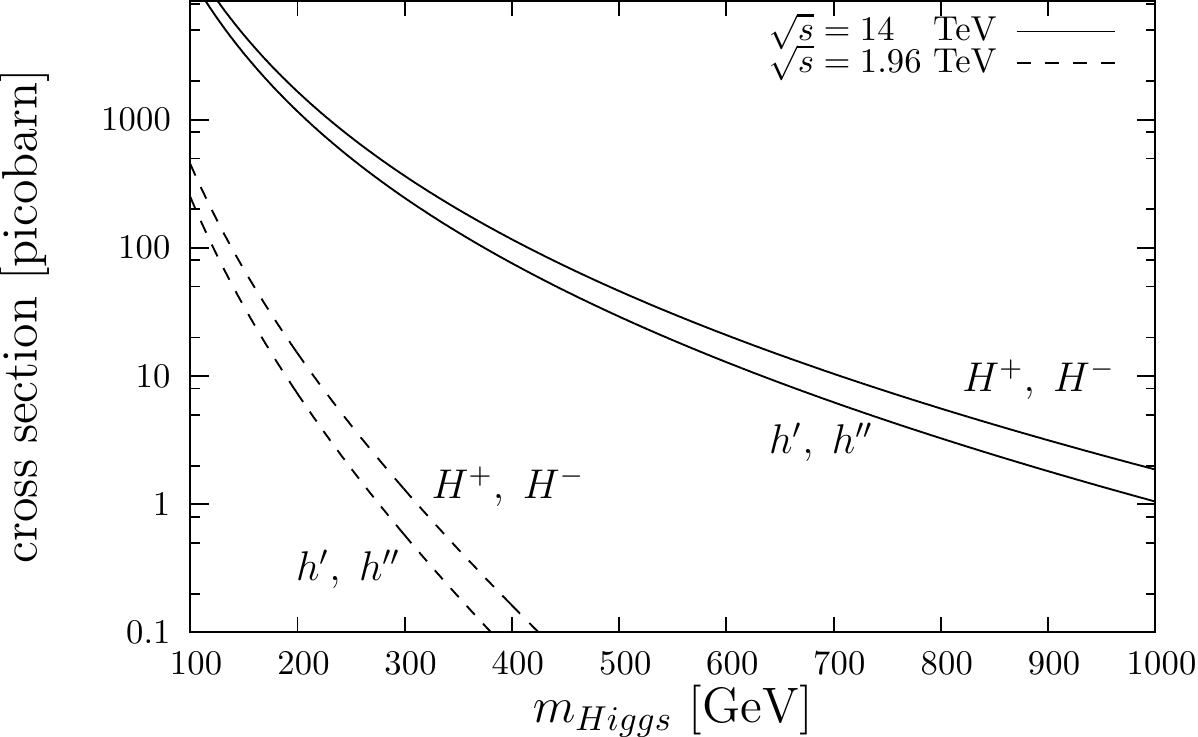}
\caption{\label{fig6}
The cross sections for Higgs-boson production via
quark--antiquark fusion  for proton--proton
collisions at LHC energies (full lines) and for proton--antiproton collisions at
Tevatron energies (dashed lines), respectively, as functions of the
Higgs-boson masses.}
\end{figure}

For the $h'$ and $h''$ we have a very large coupling proportional
to $m_t$ for $c$ quarks. For the charged Higgs bosons $H^+$ and $H^-$
there is a large coupling in the fusion processes with $c\bar{s}$
and $s\bar{c}$ quarks, respectively. There are plenty
of $c$ and $s$ quarks in the proton at LHC energies.
Thus, these processes contribute significantly to Higgs-boson
production. The total cross section for
the production of a Higgs boson $H_1$ via
$q\bar{q}'$ fusion is easily evaluated from the diagrams
of Figs.~\ref{fig4} and \ref{fig5}. We get with
$s = (p_1+p_2)^2$,
the c.m. energy squared of the process (\ref{eq50}), the
following:
\begin{multline}
\label{eq53}
\left. 
	\sigma (p (p_1) + p(p_2) \rightarrow H_1(k) + X) 
\right|_{q\bar{q}'-\text{fusion}} =\\
\frac{\pi}{3 v_0^2 s} (|a|^2+|b|^2) F_{q\bar{q}'}\bigg(\frac{m_{H_1}^2}{s} \bigg) \;.
\end{multline}
Here we define
\begin{equation}
\label{eq54}
F_{q\bar{q}'}\bigg(\frac{m_{H_1}^2}{s} \bigg) 
=
\int\limits_0^1 dx_1 N_q^p (x_1)
\int\limits_0^1 dx_2 N_{\bar{q}'}^p (x_2)
\delta \big(x_1 x_2 - \frac{m_{H_1}^2}{s} \big)
\end{equation}
where $N_q^p (x)$ and $N_{\bar{q}'}^p (x)$ are the quark
and antiquark distribution functions of the proton, respectively,
at LHC energies.
From (\ref{eq53}) and Tab. \ref{tab6} we get
for the Drell--Yan type contributions to (\ref{eq50})
\begin{multline}
\label{eq55}
\left. \sigma (p (p_1) + p(p_2) \rightarrow h' + X) \right|_{\text{DY}}
=\\
\frac{\pi}{3 v_0^2 s} \bigg[
m_t^2 F_{c\bar{c}}\bigg( \frac{m_{h'}^2}{s} \bigg) 
+
m_b^2 F_{s\bar{s}}\bigg(\frac{m_{h'}^2}{s} \bigg) \bigg] \;,
\end{multline}
\begin{multline}
\label{eq56}
\left. \sigma (p (p_1) + p(p_2) \rightarrow h'' + X) \right|_{\text{DY}}
=\\
\frac{\pi}{3 v_0^2 s} \bigg[
m_t^2 F_{c\bar{c}}\bigg(\frac{m_{h''}^2}{s} \bigg)
+
m_b^2 F_{s\bar{s}}\bigg(\frac{m_{h''}^2}{s} \bigg) \bigg] \;,
\end{multline}
\begin{multline}
\label{eq57}
\left. \sigma (p (p_1) + p(p_2) \rightarrow H^+ + X) \right|_{\text{DY}}
=\\
\frac{\pi}{3 v_0^2 s} 
(m_t^2+m_b^2)
F_{c\bar{s}}\bigg(\frac{m_{H^+}^2}{s} \bigg) \;,
\end{multline}
\begin{multline}
\label{eq58}
\left. \sigma (p (p_1) + p(p_2) \rightarrow H^- + X) \right|_{\text{DY}}
=\\
\frac{\pi}{3 v_0^2 s} 
(m_t^2+m_b^2)
F_{s\bar{c}}\bigg(\frac{m_{H^-}^2}{s} \bigg) \;.
\end{multline}
The cross sections (\ref{eq55})-(\ref{eq58})
are shown in Fig.~\ref{fig6}
for $\sqrt{s}= 14~\text{TeV}$, corresponding to the energy available at the LHC,
as function of the Higgs-boson masses.
We also show in Fig.~\ref{fig6} the
results for Higgs-boson production in proton--antiproton
collisions for $\sqrt{s}= 1.96~\text{TeV}$ 
corresponding to the energy available 
at the Tevatron. Of course, for $p\bar{p}$ collisions
the factor $F_{q\bar{q}'}$ in \eqref{eq53} and \eqref{eq54}
has to be replaced by an integral over proton and antiproton
distribution functions
\begin{multline}
\label{eq54b}
F_{q\bar{q}'}\bigg(\frac{m_{H_1}^2}{s} \bigg) 
=
\frac{1}{2}
\int\limits_0^1 dx_1 
\int\limits_0^1 dx_2 
\bigg(
N_q^p (x_1)
N_{\bar{q}'}^{\bar{p}} (x_2)\\
+
N_{{q}}^{\bar{p}} (x_1)
N_{\bar{q}'}^p (x_2)
\bigg)
\delta \big(x_1 x_2 - \frac{m_{H_1}^2}{s} \big) \,.
\end{multline}

We emphasize that all results of this subsection are obtained
in the strict symmetry limit of the MCPM.

%

%%%%%%%%%%%%%%%%%%%%%%%%%%%%%%%%%%%%%%%%%%%%%%%%%%%%%%%%%%%%%%%%%%%%%%%%
% subsection IV.2
%
%%%%%%%%%%%%%%%%%%%%%%%%%%%%%%%%%%%%%%%%%%%%%%%%%%%%%%%%%%%%%%%%%%%%%%%
\subsection{Higgs-boson production by gluon--gluon fusion}
\label{IV2}

\begin{figure}[t] 
\centering
\includegraphics[width=0.7\linewidth,clip]{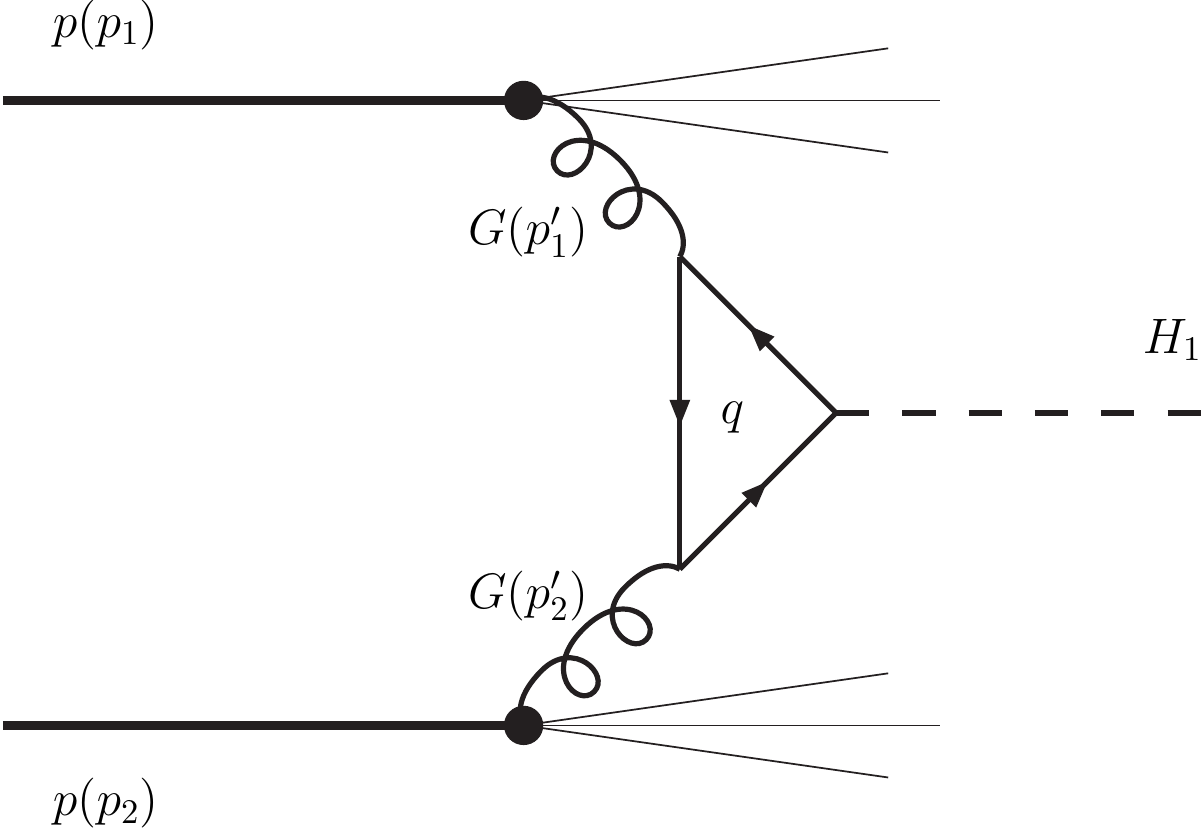}
\caption{\label{fig7}
Diagram for the production of a 
Higgs particle $H_1$ by gluon--gluon
fusion in proton--proton collisions.}
\end{figure}

Here we study the production of the neutral Higgs particles
$\rho'$, $h'$ and $h''$ via gluon--gluon fusion.
The corresponding generic diagram is shown in Fig.~\ref{fig7}.
In leading order the gluons couple to the Higgs particle
via a quark loop. The diagram of Fig.~\ref{fig7}
is easily evaluated and gives for the total cross section
\begin{multline}
\label{eq59}
\left. \sigma (p (p_1) + p(p_2) \rightarrow H_1 + X) \right|_{GG-\text{fusion}}
=\\
\frac{\pi^2 \Gamma(H_1 \rightarrow GG)}{8 \; s\; m_{H_1}}
F_{GG}\big(\frac{m_{H_1}^2}{s} \big) \;,
\end{multline}
where $H_1=\rho', h'$, and $h''$. The function $F_{GG}$
is defined as 
\begin{equation}
\label{eq60}
F_{GG}\bigg(\frac{m_{H_1}^2}{s} \bigg) 
\!=\!\!
\int\limits_0^1 dx_1 N_G^p (x_1)
\int\limits_0^1 dx_2 N_G^p (x_2)
\delta \big(x_1 x_2 -  \frac{m_{H_1}^2}{s} \big)
\end{equation}
with $N_G^p(x)$ the gluon distribution function of the proton 
at LHC energies. Furthermore, 
$\Gamma(H_1 \rightarrow GG)$ is
the partial decay width for $H_1$ decaying into
two gluons as discussed in Sect. \ref{secIII3}.

\begin{figure}[t] 
\centering
\includegraphics[width=\linewidth,clip]{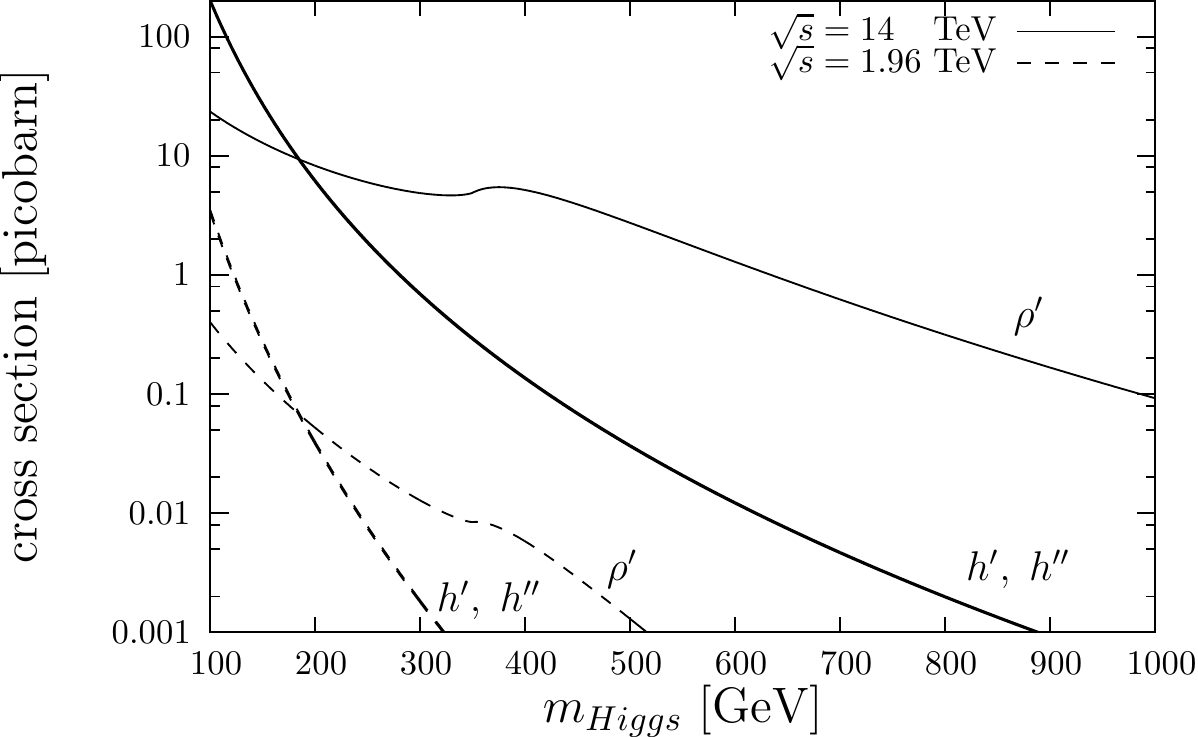}
\caption{\label{fig8}
The cross sections for the production
of $\rho'$, $h'$ and $h''$ via gluon--gluon
fusion as functions of the
Higgs-boson masses in proton--proton collisions
at a c.m. energy of $\sqrt{s}= 14~\text{TeV}$ (full lines) and
in proton--antiproton collisions at $\sqrt{s}= 1.96~\text{TeV}$ (dashed lines), respectively.} 
\end{figure}

Setting $H_1=\rho'$ in (\ref{eq59}) and using
$\Gamma(\rho' \rightarrow GG)$ from \eqref{eq37}
we get the cross section for $\rho'$ production via
gluon--gluon fusion in the MCPM. The result
as shown in Fig.~\ref{fig8} 
is valid in the strict symmetry limit of the MCPM and
coincides
with that from the SM for $\rho'_{SM}$; see for instance \cite{Djouadi:2005gi}.
Setting successively $H_1=h'$ and $H_1=h''$ in \eqref{eq59}
we obtain with \eqref{eq38} and \eqref{eq39}   
our estimates, in the sense discussed at the end 
of Sect.~\ref{secIII},
for the corresponding
production cross sections as shown in Fig.~\ref{fig8}.
Again, we give in Fig.~\ref{fig8} also the
cross sections for Higgs-boson production via gluon--gluon
fusion in proton--antiproton collisions at $\sqrt{s}=1.96$~TeV.

%%%%%%%%%%%%%%%%%%%%%%%%%%%%%%%%%%%%%%%%%%%%%%%%%%%%%%%%%%%%%%%%%%%%%%%%
% conclusion
%
%%%%%%%%%%%%%%%%%%%%%%%%%%%%%%%%%%%%%%%%%%%%%%%%%%%%%%%%%%%%%%%%%%%%%%%
\section{Discussion and conclusions}
\label{secC}

In this article we have given phenomenological predictions for
proton--proton collisions at LHC energies in the framework
of a two-Higgs-doublet model satisfying the principle of
maximal \CP\ invariance as introduced in \cite{Maniatis:2007de}.
In this maximally-CP-symmetric model (MCPM) there are three neutral Higgs particles,
$\rho'$, $h'$ and $h''$, and one charged Higgs-boson pair
$H^\pm$. We have investigated the decays of these particles.
The Higgs particle $\rho'$ behaves practically
as the Higgs particle $\rho'_{SM}$ in the SM.
Only the $2 \gamma$ widths of $\rho'$ and $\rho'_{SM}$ may
differ substantially for $m_{\rho'} \gtrsim 300$~GeV.
The particles $h'$, $h''$ and
$H^\pm$, on the other hand, are predicted to have
quite interesting properties. They couple to the
fermions of the second family with coupling constants given by
the masses of the third family. As a consequence the main
decays of these Higgs particles are the fermionic ones
of Tab.~\ref{tab4} if the Higgs masses are
below about $400~\text{GeV}$. 
For larger Higgs-boson masses, the decay into a lighter Higgs
boson associated with a gauge boson may become dominant, as
shown in examples by the branching ratios in Figs.~\ref{Hbranch}
and \ref{Hpmbranch}.

We have 
studied the production of the Higgs bosons 
$h'$ and $h''$
in proton--proton 
and proton--antiproton collisions via
quark--antiquark and gluon--gluon fusion.
We have found that the much higher gluon densities compared to the quark 
densities in the proton do
not compensate the loop suppression 
of the leading order gluon--gluon fusion process.
Thus we find
the Drell--Yan process 
with the annihilation of $c \bar{c}$ quarks
dominating the production cross sections for $h'$ and $h''$.
The Drell--Yan process also leads to a similar production
cross section for $H^+$ and $H^-$ via the
annihilation of $c \bar{s}$ and $s \bar{c}$ quarks, respectively.
In this way we get for the Higgs bosons $h'$, $h''$ and $H^\pm$,
if their masses are below $400$~GeV,
quite high production cross sections exceeding $100$~pb at LHC energies.
This is shown in Fig.~\ref{fig6}.
With an integrated luminosity of $100$~fb$^{-1}$ this
translates into the production of more than 
$10^8$ Higgs bosons
of the types $h'$, $h''$ and $H^\pm$
if their masses are around 200~GeV.
For Higgs-boson masses
of $400$~GeV the number of produced particles $h'$, $h''$ and $H^\pm$
is predicted to be of order $10^7$. 
These produced Higgs bosons will mainly decay into $c$- and $s$-quarks giving two jets.
But, of course, there is a very large background
from ordinary QCD two-jet events. Perhaps it will be possible 
to detect the Higgs-boson production events
over the QCD background using
$c$-quark tagging. Clearly, this presents an
experimental challenge.
A further possibility is to use the
information from the angular distribution of the two jets.
For the decays of the scalar particles $h'$, $h''$ and $H^\pm$ the
two-jet angular distributions must be isotropic in the rest frame
of the decaying particle. Contrary to this, the QCD two-jet
events are peaked in the beam directions. Clearly, only a
detailed Monte Carlo study 
including an investigation of the QCD background and
the detector resolution
can tell if the particles
$h'$, $h''$ and $H^\pm$ are observable in their two-jet decays with
the LHC detectors.

A promising signal for detecting the Higgs bosons $h'$, $h''$ and $H^\pm$
of the MCPM is provided by their leptonic decays
\begin{equation}
\label{eq-decmu}
\begin{split}
h'  \rightarrow& \mu^+ \mu^-\;,\\
h''  \rightarrow& \mu^+ \mu^-\;,\\
H^+  \rightarrow& \mu^+ \nu_\mu\;,\\
H^-  \rightarrow& \mu^- \bar{\nu}_\mu \;.
\end{split}
\end{equation}
In \eqref{eq-branch} we have estimated the branching
fractions for these decays to be about
$3 \times 10^{-5}$ for Higgs-boson masses below 400~GeV.
With an integrated luminosity of 100~fb$^{-1}$ and the 
number of produced Higgs bosons given above we predict
then around 3000 leptonic events for each of the channels
in \eqref{eq-decmu} if the Higgs-boson masses
are around 200~GeV. For Higgs-boson masses of
400~GeV we still have 300 leptonic events for each of
the decays in \eqref{eq-decmu}. We emphasize that a distinct
feature of the MCPM is that
decays involving the leptons $\tau$ and $\nu_\tau$ as well
as $e$ and $\nu_e$ should be highly suppressed
compared to the muonic channels \eqref{eq-decmu}. 
We may
note that the $\mu^+ \mu^-$ channel will be prominent
at the LHC for the search for new effects including for instance
heavy $Z'$ bosons or Kaluza--Klein particles, see for instance
\cite{cms:tdr,atlas:tdr}.
Thus, the suppression of the $\tau$ and $e$ channels for the
Higgs bosons of the MCPM may be an important way for
distinguishing  the MCPM from
other possibilities for physics beyond
the SM.

To conclude, we have in this article presented concrete predictions for
the production and decay of the Higgs bosons of the MCPM.
We found the Drell--Yan type process to be the dominant production mechanism.
But, of course, there are also other mechanisms, which we hope to
investigate in future work, for instance, Higgs-strahlung in 
quark--quark collisions.
Thus, the predicted numbers of produced Higgs bosons given above for the LHC
are in fact lower limits.
We are looking forward to the start up of the experimentation at the LHC, where
it should be possible to check our predictions.

%\begin{acknowledgement}
\acknowledgments
It is a pleasure for the authors to thank
P.~Braun-Munzinger, C.~Ewerz, G.~Ingelman, A.~v.Manteuffel, and H.C.~Schultz-Coulon
for useful discussions and suggestions. Thanks are due to
A.~v.Manteuffel and D.~St\"{o}ckinger for reading of the manuscript. 
%\end{acknowledgement}

\appendix
\numberwithin{equation}{section}

%%%%%%%%%%%%%%%%%%%%%%%%%%%%%%%%%%%%%%%%%%%%%%%%%%%%%%%%%%%%%%%%%%%%%%%%
% appendix A
%
%%%%%%%%%%%%%%%%%%%%%%%%%%%%%%%%%%%%%%%%%%%%%%%%%%%%%%%%%%%%%%%%%%%%%%%
\section{The Lagrangian after EWSB}
\label{appA}

The task is to express the Lagrangian $\mathscr{L}$ of the MCPM 
%$\mathscr{L}$ \eqref{eq5}
%with $\mathscr{L}_\varphi$ given in (\ref{eq6})-(\ref{eq9})
%and $\mathscr{L}_{\mathrm{Yuk}}$ given in (\ref{eq11}) in
in terms of physical fields in the unitary gauge.
This Lagrangian is given by
\begin{equation}
\label{eq5}
\mathscr{L}=\mathscr{L}_\varphi + \mathscr{L}_{\mathrm{Yuk}} + \mathscr{L}_{\mathrm{FB}}
\end{equation}
where $\mathscr{L}_{\mathrm{FB}}$ is the standard gauge kinetic
term for the fermions and gauge bosons; see for instance 
\cite{Nachtmann:1990ta}. The Higgs-boson Lagrangian
is denoted by $\mathscr{L}_\varphi$, the Yukawa term, giving
the coupling of the fermions to the Higgs fields, by
$\mathscr{L}_{\mathrm{Yuk}}$. 
In \cite{Maniatis:2007vn,Maniatis:2007de}
the form for $\mathscr{L}_\varphi$ and
$\mathscr{L}_{\mathrm{Yuk}}$ was derived from the
requirement of maximal CP invariance, absence of flavor-changing
neutral currents and absence of mass-degenerate massive fermions. For $\mathscr{L}_\varphi$
the result is
\begin{equation}
\label{eq6}
\mathscr{L}_\varphi =
\sum\limits_{i=1,2}
\big( D_\mu \varphi_i \big)^\dagger
\big( D^\mu \varphi_i \big) 
- V( \varphi_1, \varphi_2) \;,
\end{equation}
where $D_\mu$ are the covariant derivatives and $V$
is given in~\eqref{eq8}. The Yukawa term,
$\mathscr{L}_{\mathrm{Yuk}}$, is given in~\eqref{eq11}.

Using the unitary gauge we insert for the
Higgs-boson fields $\varphi_1$ and $\varphi_2$
the expressions~\eqref{eq17} and \eqref{eq18}, respectively.
In the following we use
as independent parameters of the Lagrangian 
the fine structure constant $\alpha$, respectively
$e=\sqrt{4 \pi \alpha}$, the Fermi constant $G_F$,
the mass $m_Z$ of the Z-boson, the
Higgs-boson masses
$m_{\rho'}^2$, 
$m_{h'}^2$, 
$m_{h''}^2$, 
$m_{H^\pm}^2$, see (\ref{eq19})-(\ref{eq22}),
and the fermion masses $m_\tau$, $m_t$, $m_b$; see (\ref{eq24}).
With this, the following parameters are dependent ones:
$s_W \equiv \sin \theta_W$,
$c_W \equiv \cos \theta_W$,
where $\theta_W$ is the weak mixing angle,
the mass $m_W$ of the W boson,
and the VEV $v_0$. The corresponding
tree-level expressions for them in terms
of the independent parameters are
\begin{equation}
\begin{split}
s_W^2 &= \frac{1}{2} \bigg[1 - (1- \frac{e^2}{\sqrt{2} G_F m_Z^2})^{1/2} \bigg]\,,\\
m_W^2 &= \frac{m_Z^2}{2} \bigg[1 + (1- \frac{e^2}{\sqrt{2} G_F m_Z^2})^{1/2} \bigg]\,,\\
v_0 &= 2^{-1/4} G_F^{-1/2}\,.
\end{split}
\end{equation}
Keeping this in mind we find
for $K_0$--$K_3$ of (\ref{eq-kdef}),
inserting (\ref{eq17}) and (\ref{eq18}),
\begin{equation}
\begin{split}
K_0 =& \frac{1}{2} (v_0 + \rho')^2
+ \frac{1}{2} (h'^2 + h''^2)
+ H^+ H^- \;,\\
K_1 =& (v_0 + \rho') h'\;,\\
K_2 =& (v_0 + \rho') h''\;,\\
K_3 =& \frac{1}{2}(v_0 + \rho')^2 - \frac{1}{2} (h'^2 + h''^2)
- H^+ H^- \;.
\end{split}
\end{equation}

We get from (\ref{eq5}) the following explicit form
of $\mathscr{L}$. The expression for the
fermion--boson term $\mathscr{L}_{\mathrm{FB}}$ is
standard and can be found for instance in \cite{Nachtmann:1990ta}.
For $\mathscr{L}_\varphi + \mathscr{L}_{\mathrm{Yuk}}$ we get
\begin{multline*}
\mathscr{L}_\varphi + \mathscr{L}_{\mathrm{Yuk}} =\hfill
\end{multline*}
\vspace{-1cm}
\begin{multline*}
\phantom{\;+\;}\frac{1}{8} m_{\rho'}^2 v_0^2 \; 
+ \frac{1}{2} (\partial_\mu \rho') (\partial^\mu \rho')\hfill
\end{multline*}
\vspace{-1.0cm}
\begin{multline*}
+ m_W^2 W_\mu^- W^{+ \, \mu} (1+ \frac{\rho'}{v_0})^2\hfill
\end{multline*}
\vspace{-1.0cm}
\begin{multline*}
+ \frac{1}{2} m_Z^2 Z_\mu Z^{\mu} (1+ \frac{\rho'}{v_0})^2 
+ (\partial_\mu H^+) (\partial^\mu H^-)\hfill
\end{multline*}
\vspace{-0.75cm}
\begin{multline*}
+ \frac{1}{2} (\partial_\mu h') (\partial^\mu h')
+ \frac{1}{2} (\partial_\mu h'') (\partial^\mu h'')\\
\end{multline*}
\vspace{-1.5cm}
\begin{multline*}
- \frac{1}{2} m_{\rho'}^2 
	\big(
	\rho'^2 + \frac{1}{v_0} \rho'^3 + \frac{1}{4 v_0^2} \rho'^4 
	\big)\\
\end{multline*}
\vspace{-1.5cm}
\begin{multline*}	
- \frac{1}{2} m_{h'}^2 h'^2 - \frac{1}{2} ( m_{\rho'}^2 + 2 m_{h'}^2 ) h'^2
	\left[ \frac{1}{v_0} \rho' + \frac{1}{2 v_0^2} \rho'^2 \right] \\
\end{multline*}
\vspace{-1.5cm}
\begin{multline*}
- \frac{1}{2} m_{h''}^2 h''^2 -  \frac{1}{2} ( m_{\rho'}^2 + 2 m_{h''}^2 ) h''^2
	\left[ \frac{1}{v_0} \rho' + \frac{1}{2 v_0^2} \rho'^2 \right] \\
\end{multline*}
\vspace{-1.5cm}
\begin{multline*}
- m_{H^\pm}^2 H^+ H^- - ( m_{\rho'}^2 + 2 m_{H^\pm}^2 ) H^+ H^-
	\left[ \frac{1}{v_0} \rho' + \frac{1}{2 v_0^2} \rho'^2 \right] \\
\end{multline*}
\vspace{-1.5cm}
\begin{multline*}
- \frac{m_{\rho'}^2}{2 v_0^2}
	\bigg[
		\frac{1}{4} ( h'^4 + h''^4 + 2 h'^2 h''^2 )\\
		+(h'^2 + h''^2) H^+ H^- + (H^+)^2 (H^-)^2
	\bigg]\\
\end{multline*}
\vspace{-1.5cm}
\begin{multline*}
+ i e \big( \frac{c_W^2-s_W^2}{2 s_W c_W} Z^\mu + A^\mu \big)
	\big( H^+ \partial_\mu H^- - H^- \partial_\mu H^+ \big)\\
\end{multline*}
\vspace{-1.5cm}
\begin{multline*}
+  \frac{e}{2 s_W c_W} Z^\mu
	\big( h'' \partial_\mu h' - h' \partial_\mu h'' \big)\\
\end{multline*}
\vspace{-1.5cm}
\begin{multline*}
+  \frac{i e}{2 s_W} W^{+\, \mu}
	\big( h' \partial_\mu H^- - H^- \partial_\mu h' \big)\\
\end{multline*}
\vspace{-1.5cm}
\begin{multline*}
-  \frac{i e}{2 s_W} W^{-\, \mu}
	\big( h' \partial_\mu H^+ - H^+ \partial_\mu h' \big)\\
\end{multline*}
\vspace{-1.5cm}
\begin{multline*}
-  \frac{e}{2 s_W} W^{+\, \mu}
	\big( h'' \partial_\mu H^- - H^- \partial_\mu h'' \big)\\
\end{multline*}
\vspace{-1.5cm}
\begin{multline*}
-  \frac{e}{2 s_W} W^{-\, \mu}
	\big( h'' \partial_\mu H^+ - H^+ \partial_\mu h'' \big)\\
\end{multline*}
\vspace{-1.5cm}
\begin{multline*}
+ e^2 \bigg[ \left(\frac{c_W^2-s_W^2}{2 s_W c_W}\right)^2  Z_\mu Z^\mu H^+ H^-\\
			+ \frac{c_W^2-s_W^2}{s_W c_W}  Z_\mu A^\mu H^+ H^-
			+ A_\mu A^\mu H^+ H^- \bigg] \\
\end{multline*}
\vspace{-1.5cm}
\begin{multline*}
 + \frac{e^2}{8 s_W^2 c_W^2} Z_\mu Z^\mu (h'^2 + h''^2 )\\
\end{multline*}
\vspace{-1.5cm}
\begin{multline*}
 + \frac{e^2}{2 s_W^2} W_\mu^+ W^{- \, \mu} 
		\big( \frac{1}{2} h'^2 + \frac{1}{2} h''^2 + H^+ H^- \big)\\
\end{multline*}
\vspace{-1.5cm}
\begin{multline*}
 + \frac{e^2}{2 s_W} \big(-\frac{s_W}{c_W} Z_\mu + A_\mu \big) W^{+\, \mu} H^- h'\\
\end{multline*}
\vspace{-1.5cm}
\begin{multline*}
 + \frac{e^2}{2 s_W} \big(-\frac{s_W}{c_W} Z_\mu + A_\mu \big) W^{-\, \mu} H^+ h'\\
\end{multline*}
\vspace{-1.5cm}
\begin{multline*}
 + \frac{i e^2}{2 s_W} \big(-\frac{s_W}{c_W} Z_\mu + A_\mu \big) W^{+\, \mu} H^- h''\\
\end{multline*}
\vspace{-1.5cm}
\begin{multline*}
 - \frac{i e^2}{2 s_W} \big(-\frac{s_W}{c_W} Z_\mu + A_\mu \big) W^{-\, \mu} H^+ h''\\
\end{multline*}
\vspace{-1.5cm}
\begin{multline*}
 - m_\tau   \bar{\tau} \tau \big( 1+ \frac{1}{v_0} \rho' \big) \\
\end{multline*}
\vspace{-1.5cm}
\begin{multline*}
 - m_t  \bar{t} t \big( 1+ \frac{1}{v_0} \rho' \big) \\
\end{multline*}
\vspace{-1.5cm}
\begin{multline*}
 - m_b  \bar{b} b \big( 1+ \frac{1}{v_0} \rho' \big) \\
\end{multline*}
\vspace{-1.5cm}
\begin{multline*}
 + \frac{m_\tau}{v_0} \bar{\mu} \mu h'
+ \frac{m_t}{v_0} \bar{c} c h'
+ \frac{m_b}{v_0} \bar{s} s h'\\
\end{multline*}
\vspace{-1.5cm}
\begin{multline*}
 + \frac{i m_\tau}{v_0} \bar{\mu} \gamma_5 \mu h''
  - \frac{i m_t}{v_0} \bar{c} \gamma_5 c h''
  + \frac{i m_b}{v_0} \bar{s} \gamma_5 s h'' \\
\end{multline*}
\vspace{-1.5cm}
\begin{multline*}
 + \frac{m_\tau}{\sqrt{2} v_0} 
		\left[
			\bar{\nu}_\mu (1+\gamma_5) \mu H^+ + \bar{\mu} (1-\gamma_5) \nu_\mu H^- 
		\right]\\
\end{multline*}
\vspace{-1.5cm}
\begin{multline}
\label{eqA2}
 - \frac{1}{\sqrt{2} v_0} 
		\bigg\{
			\bar{c}\; \big[m_t (1-\gamma_5) - m_b (1+\gamma_5) \big]\; s  H^+ \\
			+ \bar{s}\; \big[m_t (1+\gamma_5) - m_b (1-\gamma_5) \big]\; c  H^- 
		\bigg\} .\\
\end{multline}
From (\ref{eqA2}) it is easy to read off the Feynman rules
in the unitary gauge. We list here only the Higgs-boson--fermion
vertices and the vertices for two Higgs bosons and one gauge boson.
The arrow on the $W$ and $H$ lines indicates the flow of negative charge.
In case a momentum occurs in the Feynman rules, the
momentum direction is indicated by an extra arrow.\\
% tau tau rho'
%
\begin{tabular}{m{0.5\linewidth}m{0.4\linewidth}}
\includegraphics[width=.7\linewidth]{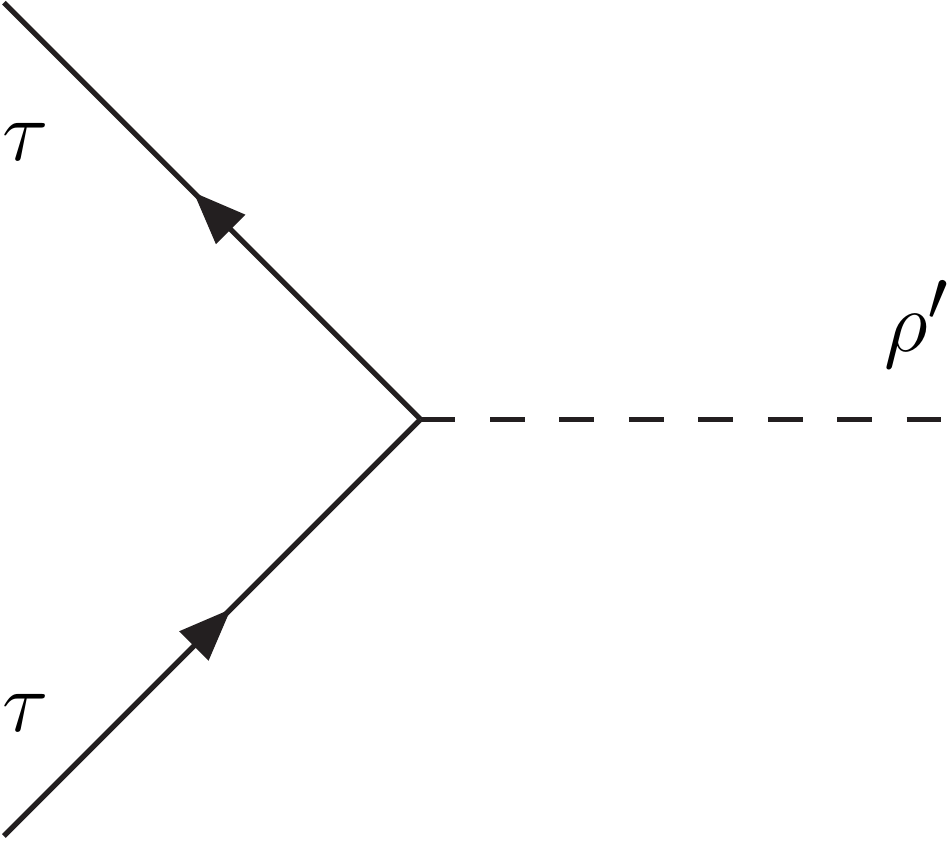} 
\hspace{0.1\linewidth}
&
{\large $-i \frac{m_\tau}{v_0}$}
\\
\end{tabular}\\
%
% t t rho'
%
\begin{tabular}{m{0.5\linewidth}m{0.4\linewidth}}
\includegraphics[width=.7\linewidth]{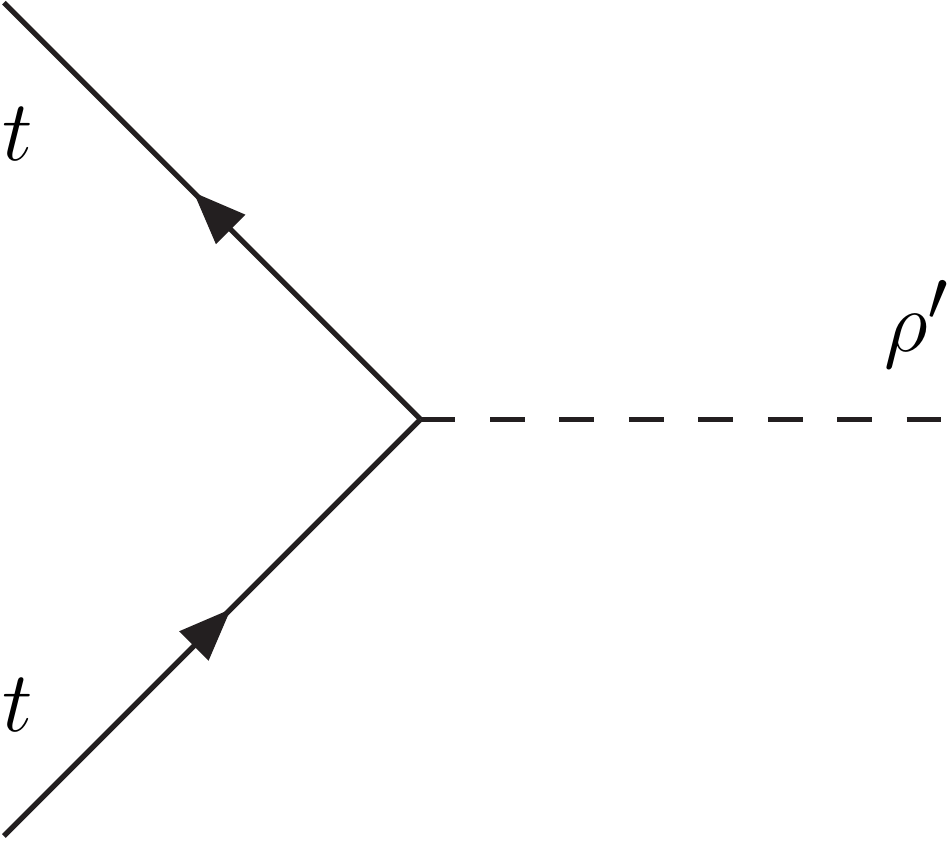} 
\hspace{0.1\linewidth}
&
{\large $-i \frac{m_t}{v_0}$}
\\
\end{tabular}\\
%
% b b rho'
%
\begin{tabular}{m{0.5\linewidth}m{0.4\linewidth}}
\includegraphics[width=.7\linewidth]{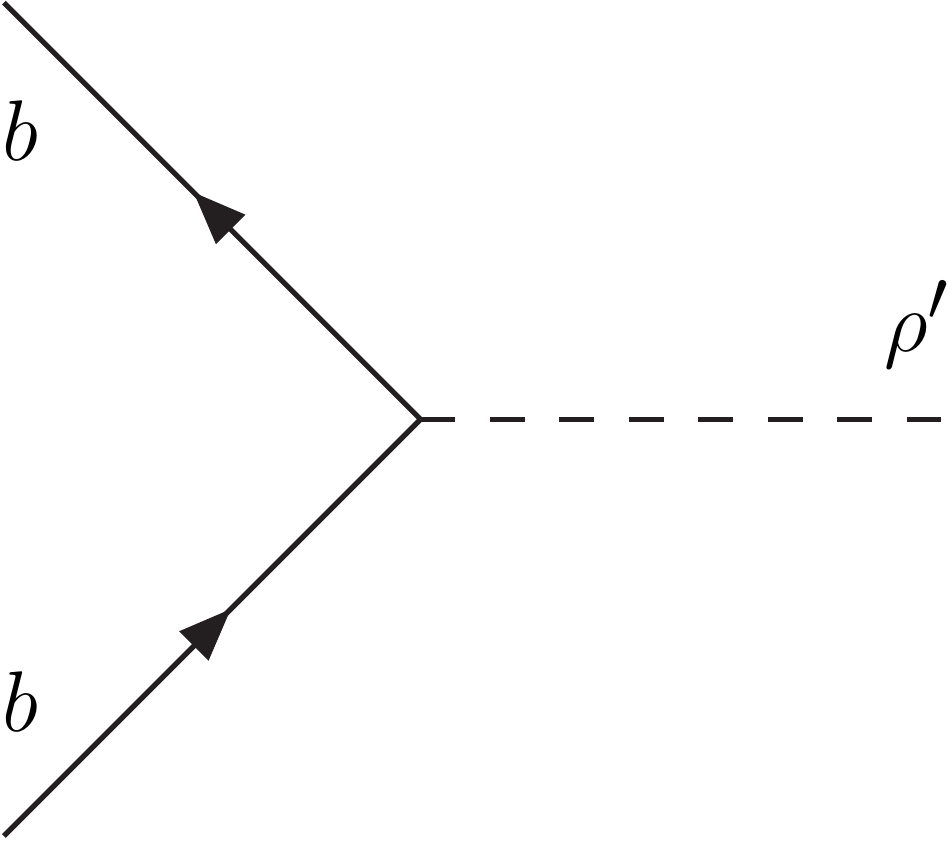} 
\hspace{0.1\linewidth}
&
{\large $-i \frac{m_b}{v_0}$}
\\
\end{tabular}\\
%
% mu mu h'
%
\begin{tabular}{m{0.5\linewidth}m{0.4\linewidth}}
\includegraphics[width=.7\linewidth]{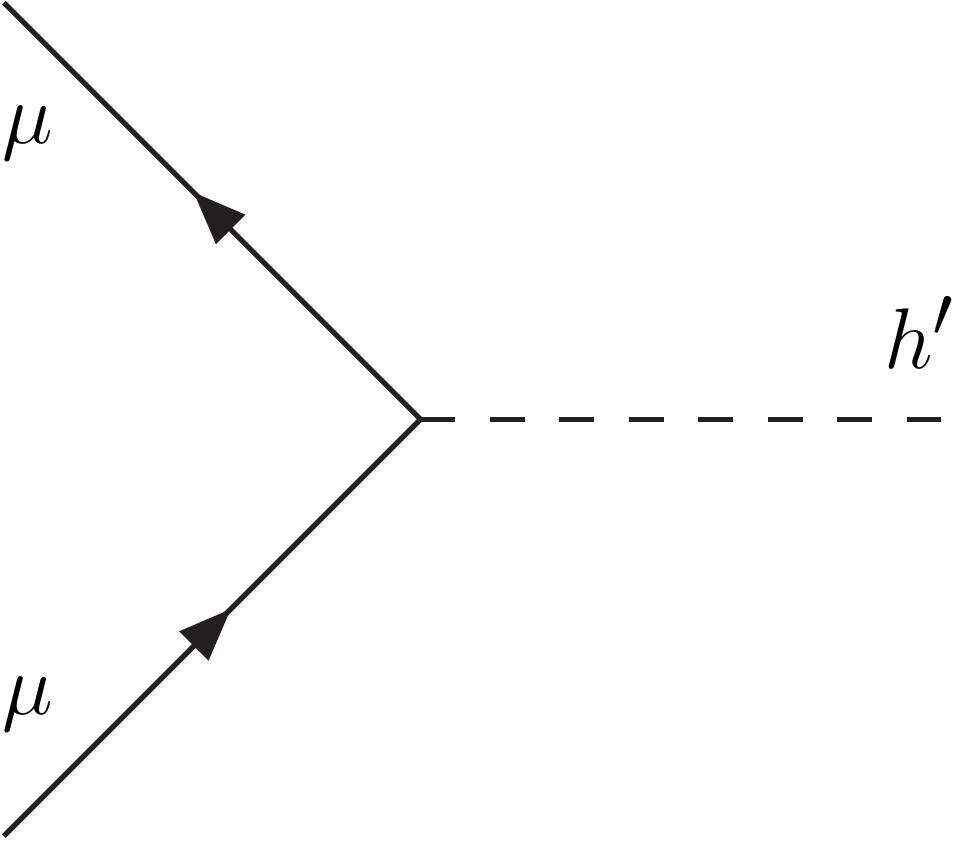} 
\hspace{0.1\linewidth}
&
{\large $ i \frac{m_\tau}{v_0}$}
\\
\end{tabular}\\
%
% c c h'
%
\begin{tabular}{m{0.5\linewidth}m{0.4\linewidth}}
\includegraphics[width=.7\linewidth]{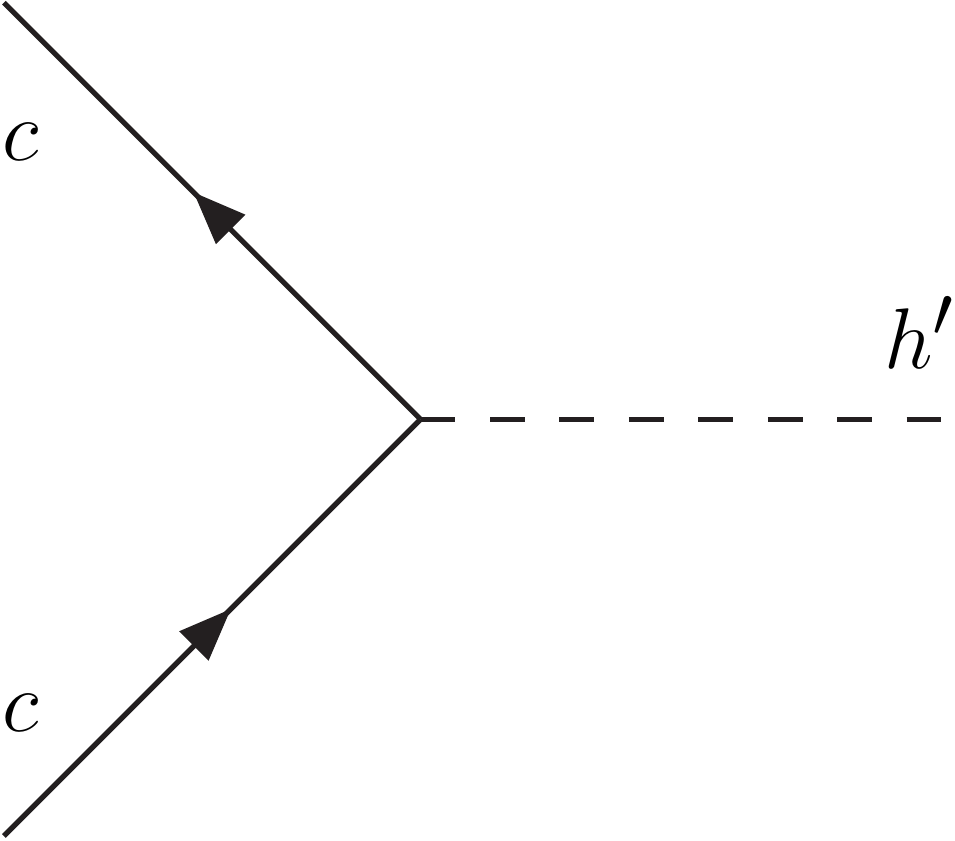} 
\hspace{0.1\linewidth}
&
{\large $ i \frac{m_t}{v_0}$}
\\
\end{tabular}\\
%
% s s h'
%
\begin{tabular}{m{0.5\linewidth}m{0.4\linewidth}}
\includegraphics[width=.7\linewidth]{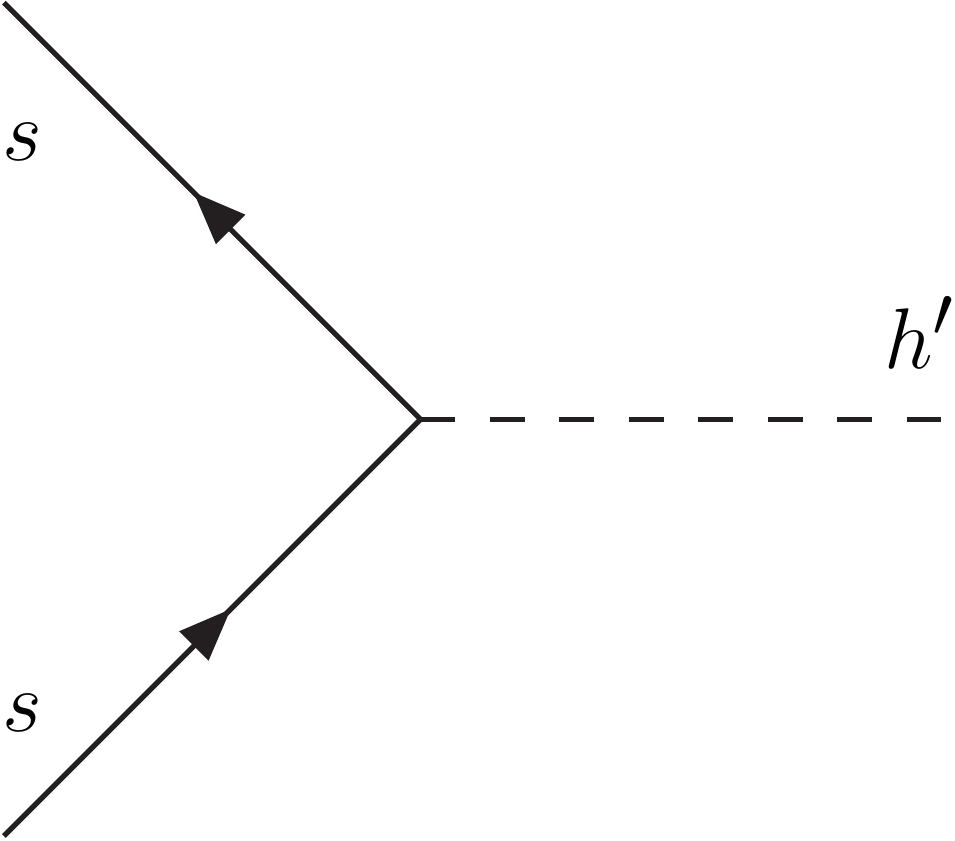} 
\hspace{0.1\linewidth}
&
{\large $ i \frac{m_b}{v_0}$}
\\
\end{tabular}\\
%
% mu mu h''
%
\begin{tabular}{m{0.5\linewidth}m{0.4\linewidth}}
\includegraphics[width=.7\linewidth]{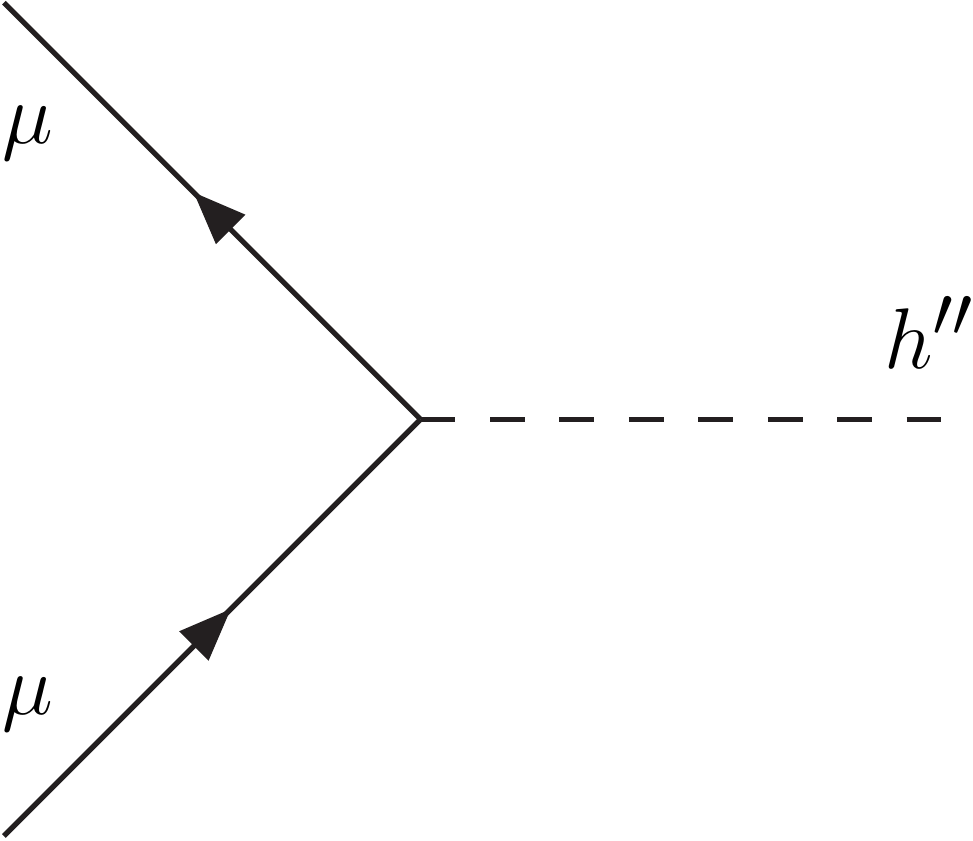} 
\hspace{0.1\linewidth}
&
{\large $ -\frac{m_\tau}{v_0} \gamma_5$}
\\
\end{tabular}\\
%
% c c h''
%
\begin{tabular}{m{0.5\linewidth}m{0.4\linewidth}}
\includegraphics[width=.7\linewidth]{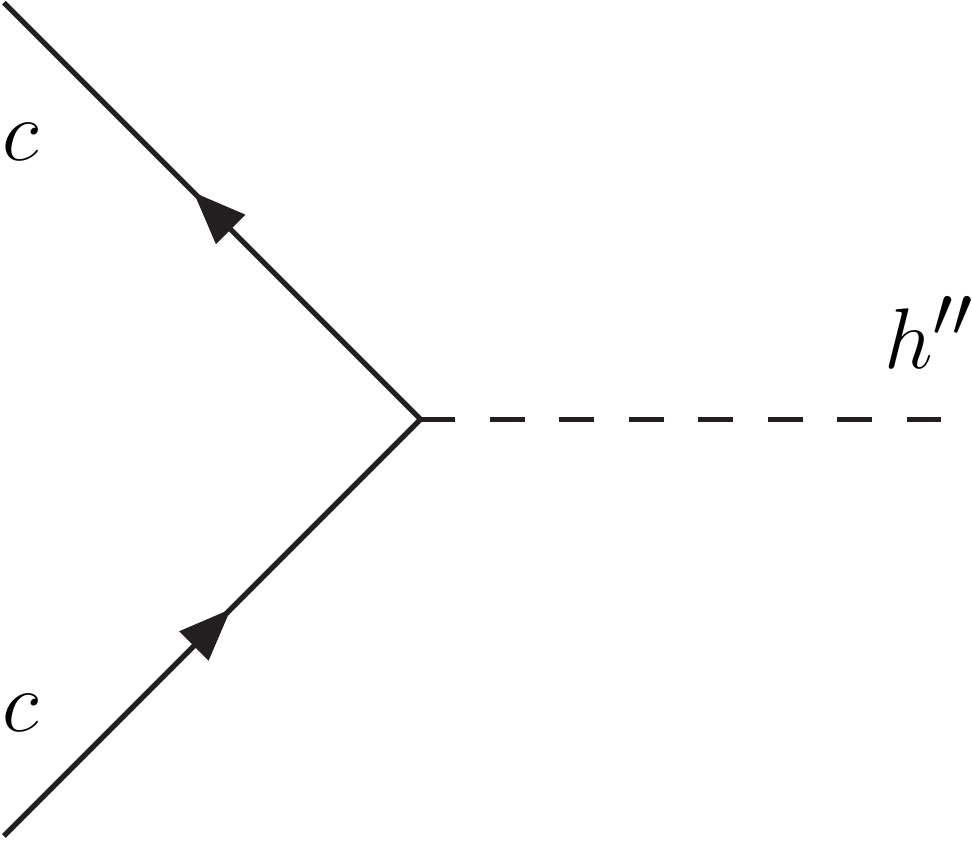} 
\hspace{0.1\linewidth}
&
{\large $ \frac{m_t}{v_0} \gamma_5$}
\\
\end{tabular}\\
%
% s s h''
%
\begin{tabular}{m{0.5\linewidth}m{0.4\linewidth}}
\includegraphics[width=.7\linewidth]{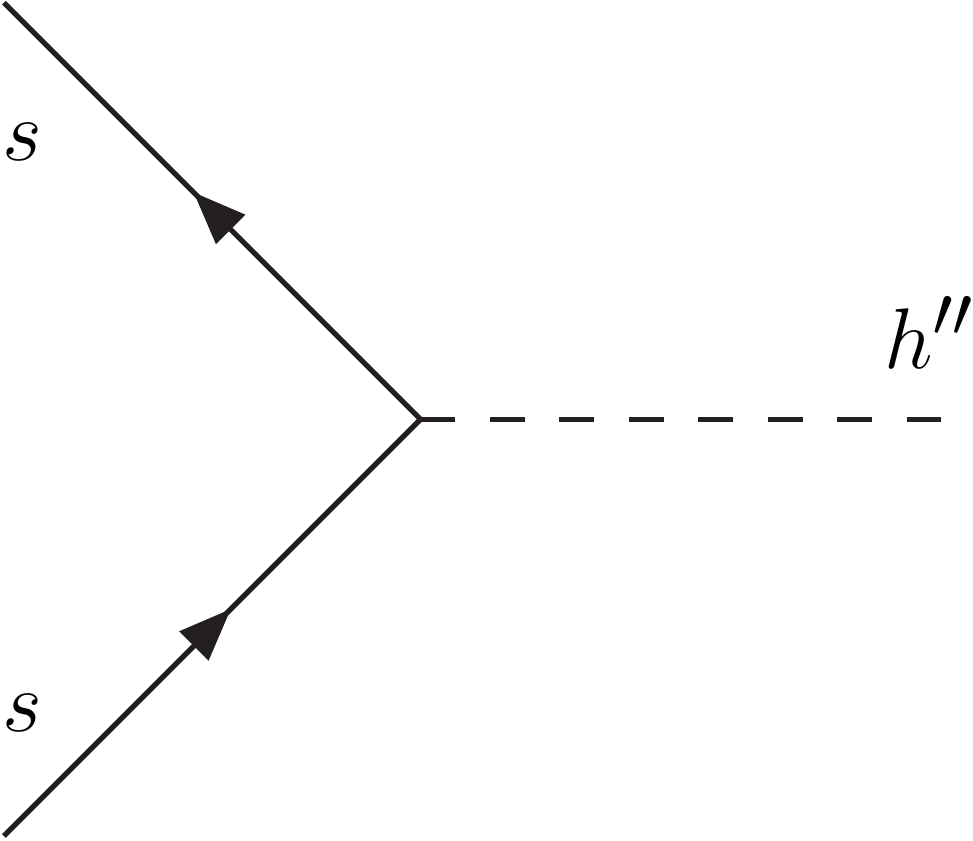} 
\hspace{0.1\linewidth}
&
{\large $ -\frac{m_b}{v_0} \gamma_5$}
\\
\end{tabular}\\
%
% nu mu H
%
\begin{tabular}{m{0.5\linewidth}m{0.4\linewidth}}
\includegraphics[width=.7\linewidth]{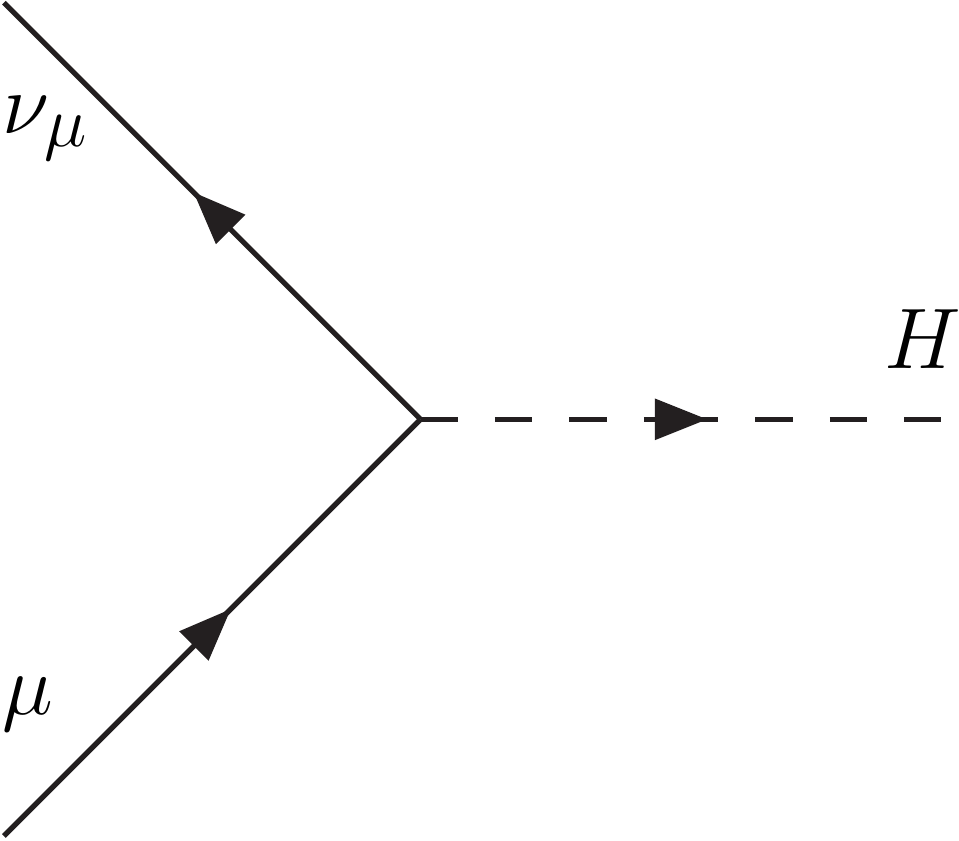} 
\hspace{0.1\linewidth}
&
{\large $ i \frac{m_\tau}{\sqrt{2} v_0} (1+\gamma_5)$}
\\
\end{tabular}\\
%
% mu nu H
%
\begin{tabular}{m{0.5\linewidth}m{0.4\linewidth}}
\includegraphics[width=.7\linewidth]{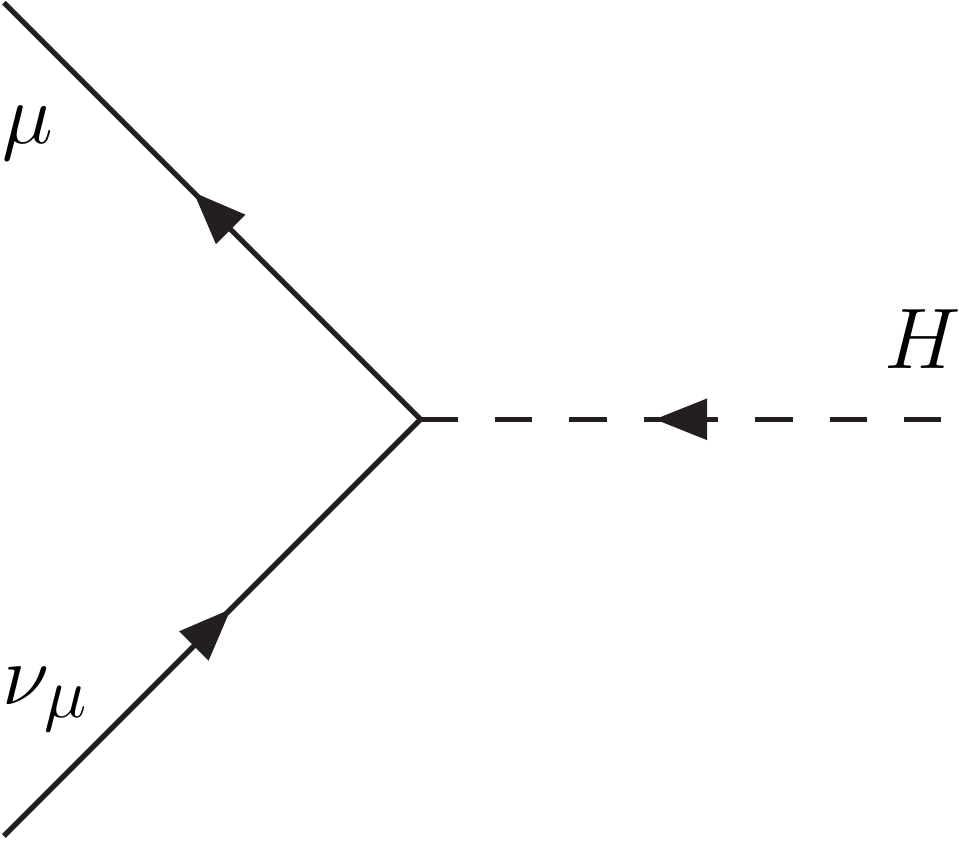} 
\hspace{0.1\linewidth}
&
{\large $ i \frac{m_\tau}{\sqrt{2} v_0} (1-\gamma_5)$}
\\
\end{tabular}\\
%
% c s H
%
\begin{tabular}{m{0.5\linewidth}m{0.4\linewidth}}
\includegraphics[width=.7\linewidth]{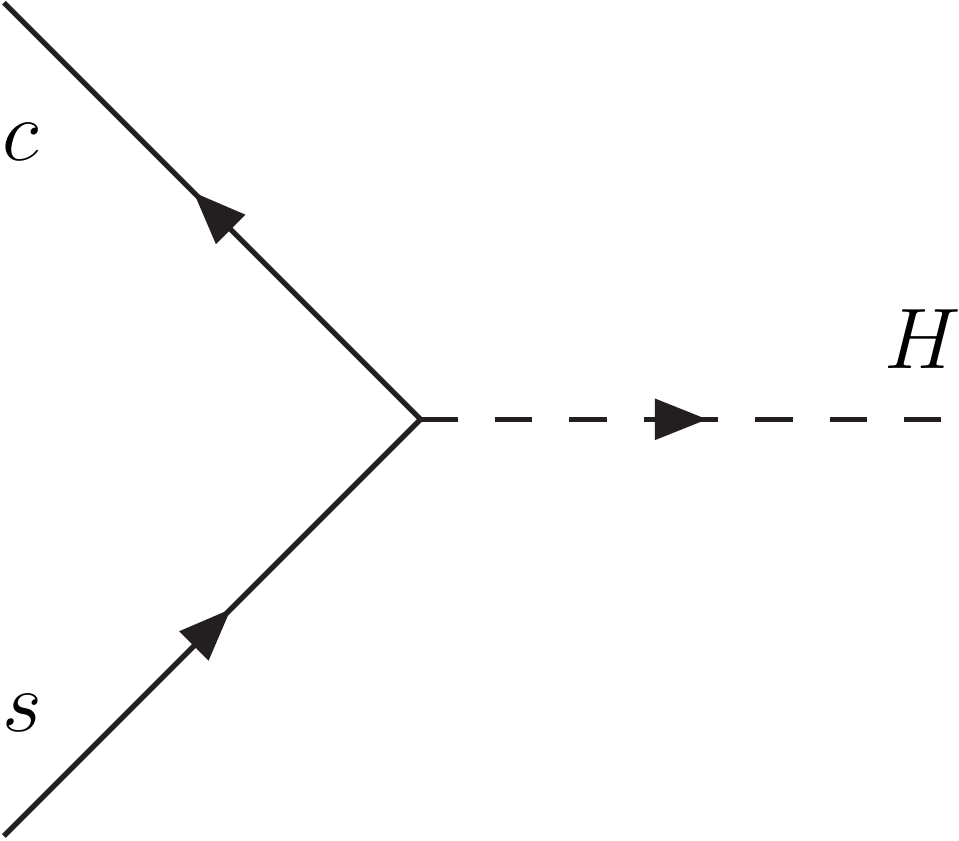} 
\hspace{0.1\linewidth}
&
{\large $ - i \frac{1}{\sqrt{2} v_0} 
	\big[
	 m_t (1-\gamma_5)
	-m_b (1+\gamma_5)
	\big]
$}
\\
\end{tabular}\\
%
% c s H 2
%
\begin{tabular}{m{0.5\linewidth}m{0.4\linewidth}}
\includegraphics[width=.7\linewidth]{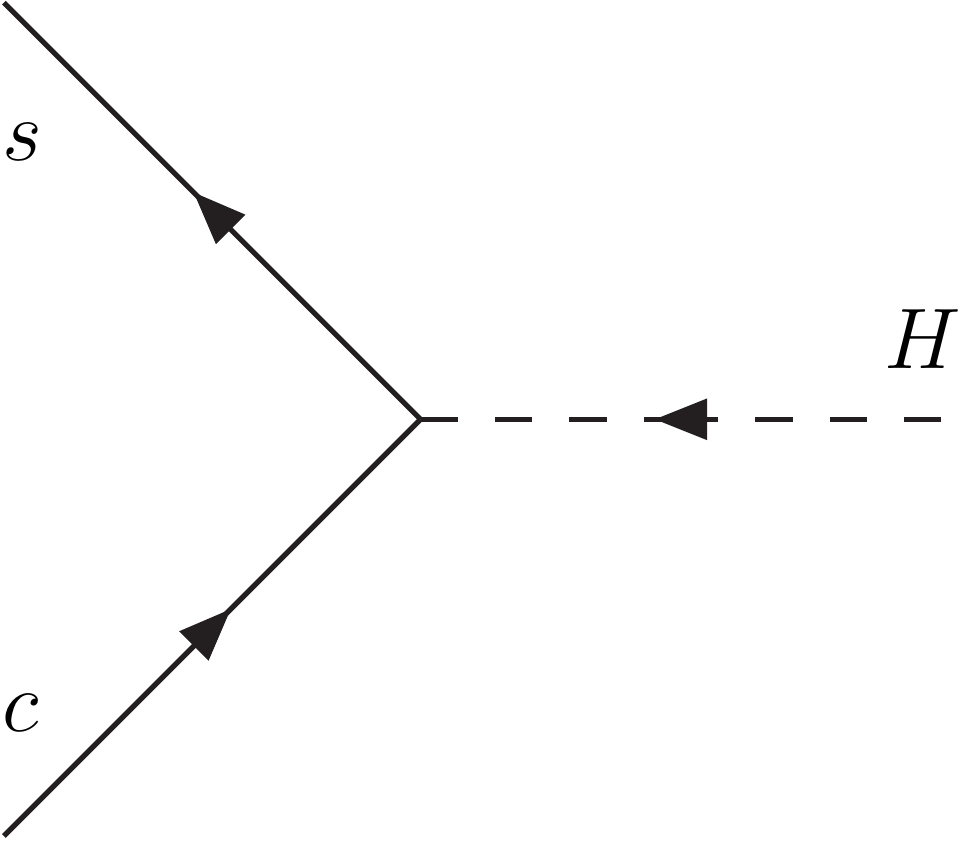} 
\hspace{0.1\linewidth}
&
{\large $ - i \frac{1}{\sqrt{2} v_0} 
	\big[
	 m_t (1+\gamma_5)
	-m_b (1-\gamma_5)
	\big]
$}
\\
\end{tabular}\\
%
% H H A 
%
\begin{tabular}{m{0.5\linewidth}m{0.4\linewidth}}
\includegraphics[width=.7\linewidth]{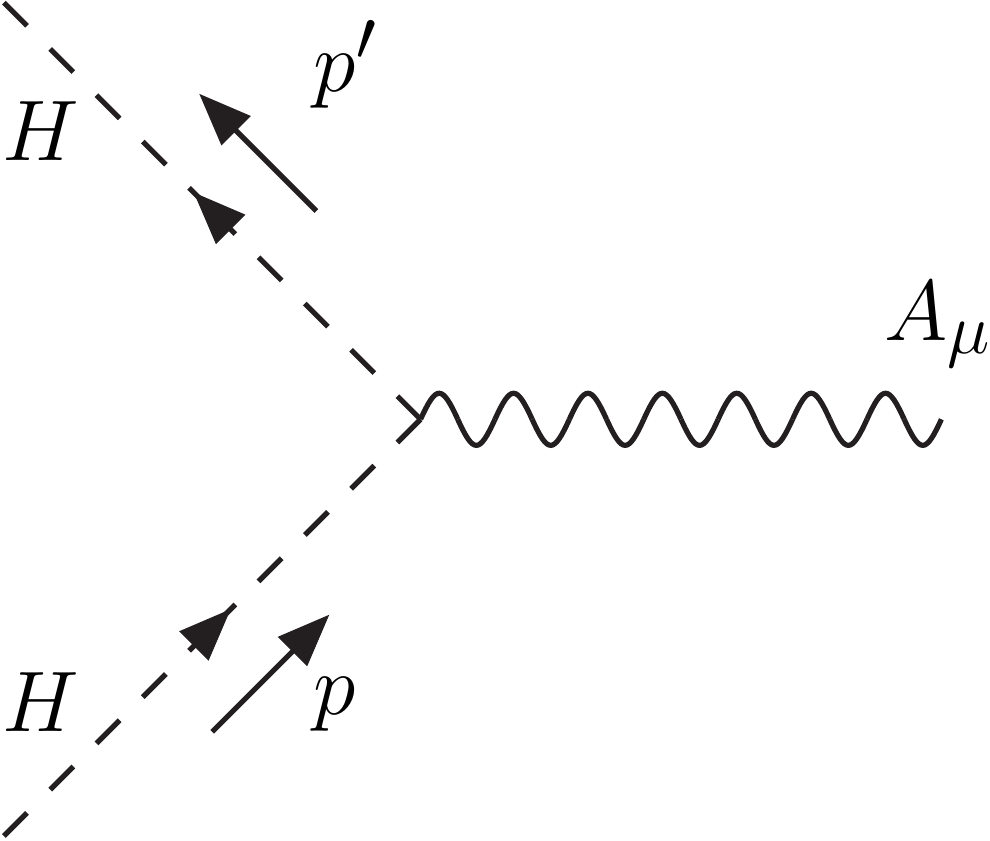} 
\hspace{0.1\linewidth}
&
{\large $ i e
	(p + p' )_\mu
$}
\\
\end{tabular}\\
%
% H H Z
%
\begin{tabular}{m{0.5\linewidth}m{0.4\linewidth}}
\includegraphics[width=.7\linewidth]{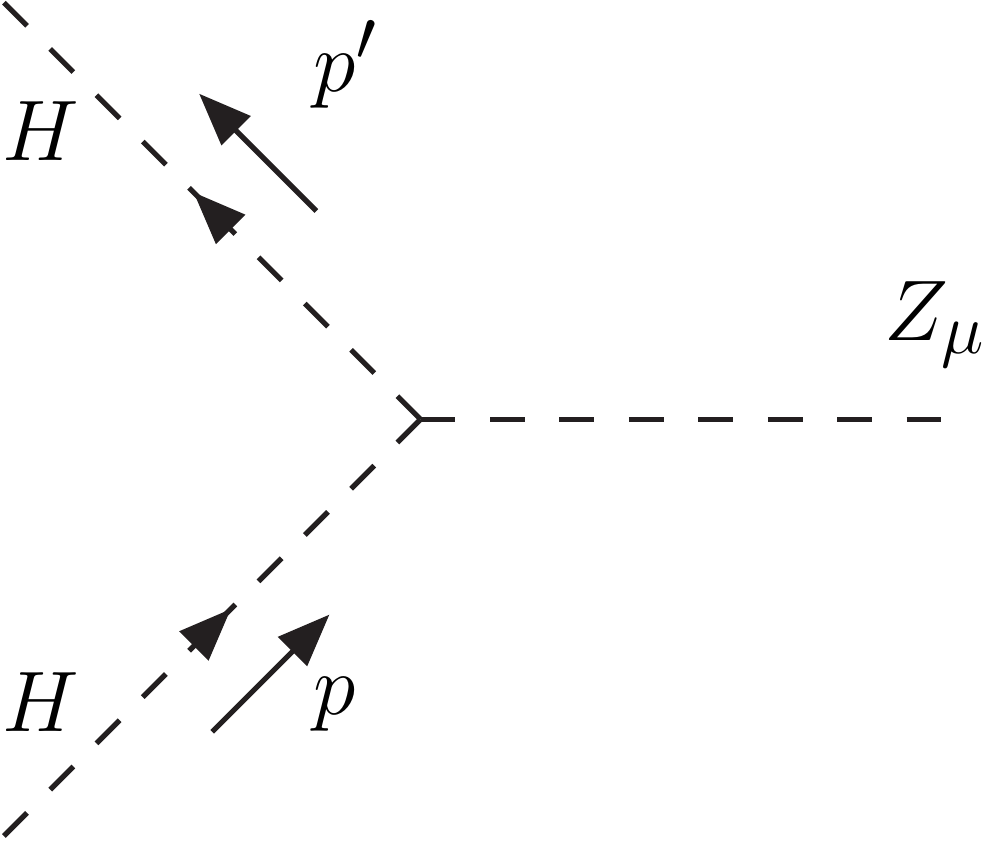} 
\hspace{0.1\linewidth}
&
{\large $ i e 
\frac{c_W^2-s_W^2}{2 s_W c_W}(p + p')_\mu
$}
\\
\end{tabular}\\
%
% h' h'' Z 
%
\begin{tabular}{m{0.5\linewidth}m{0.4\linewidth}}
\includegraphics[width=.7\linewidth]{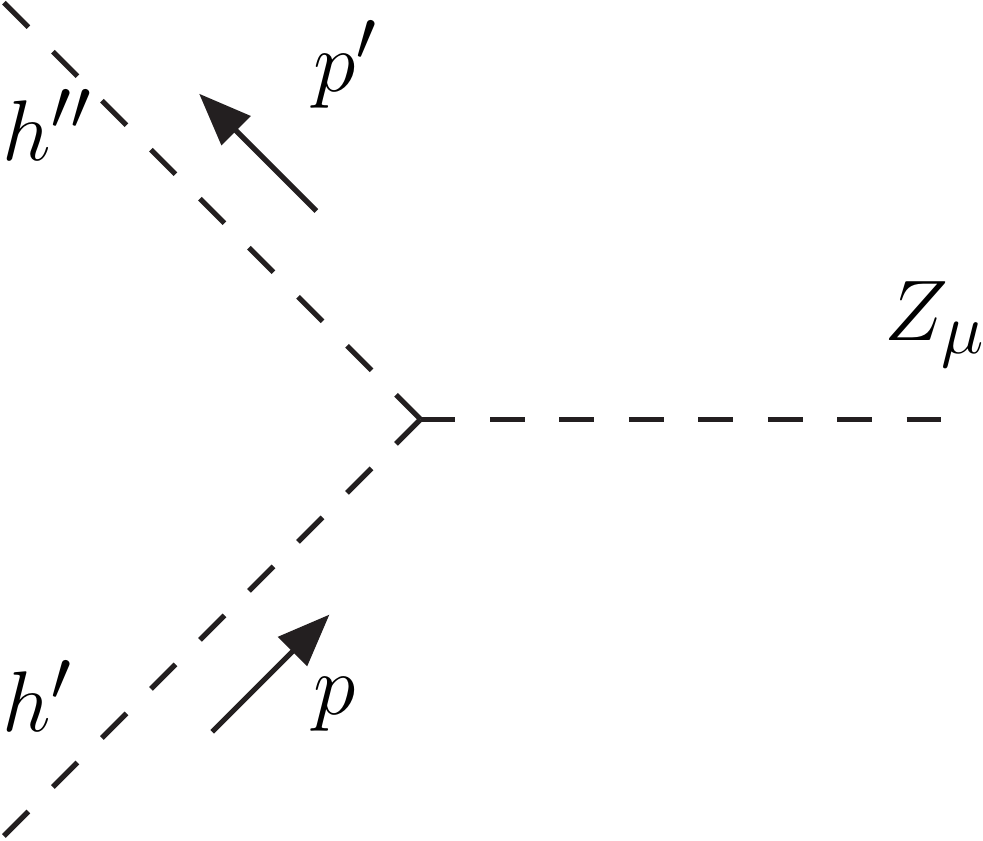} 
\hspace{0.1\linewidth}
&
{\large $ \frac{e}{2 s_W c_W}
	(p + p' )_\mu
$}
\\
\end{tabular}\\
%
% h' H W 
%
\begin{tabular}{m{0.5\linewidth}m{0.4\linewidth}}
\includegraphics[width=.7\linewidth]{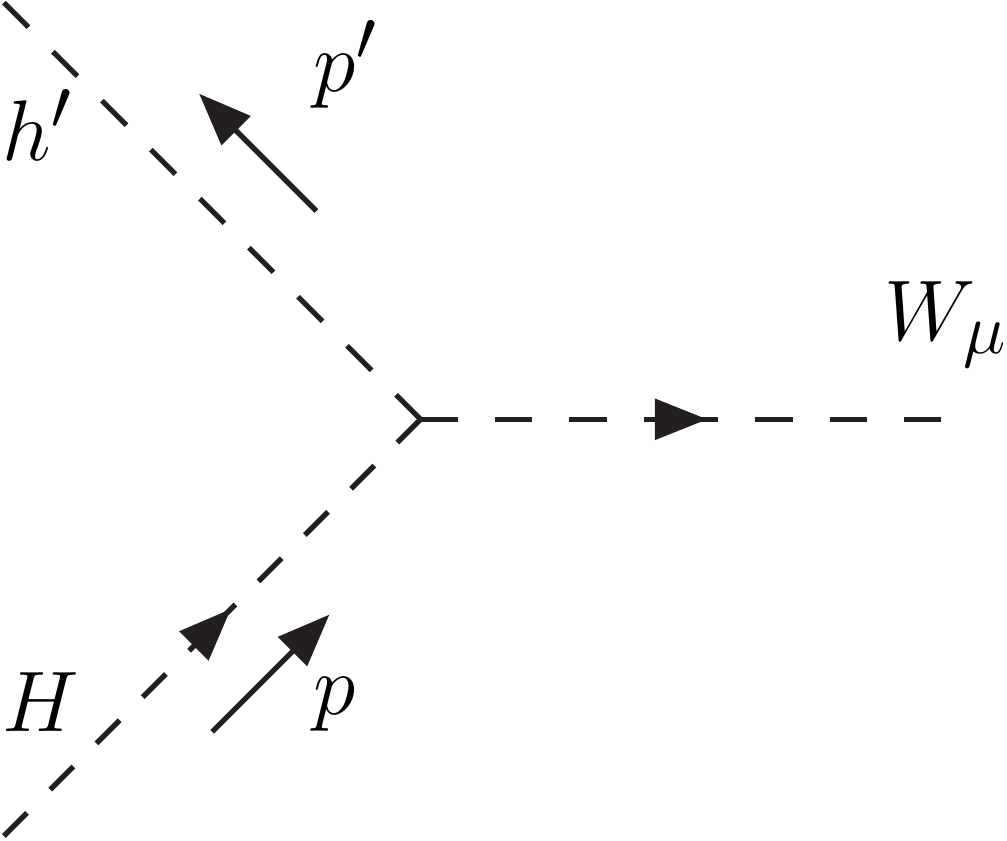} 
\hspace{0.1\linewidth}
&
{\large $ i \frac{e}{2 s_W}
	(p + p' )_\mu
$}
\\
\end{tabular}\\
%
% h' H W 2
%
\begin{tabular}{m{0.5\linewidth}m{0.4\linewidth}}
\includegraphics[width=.7\linewidth]{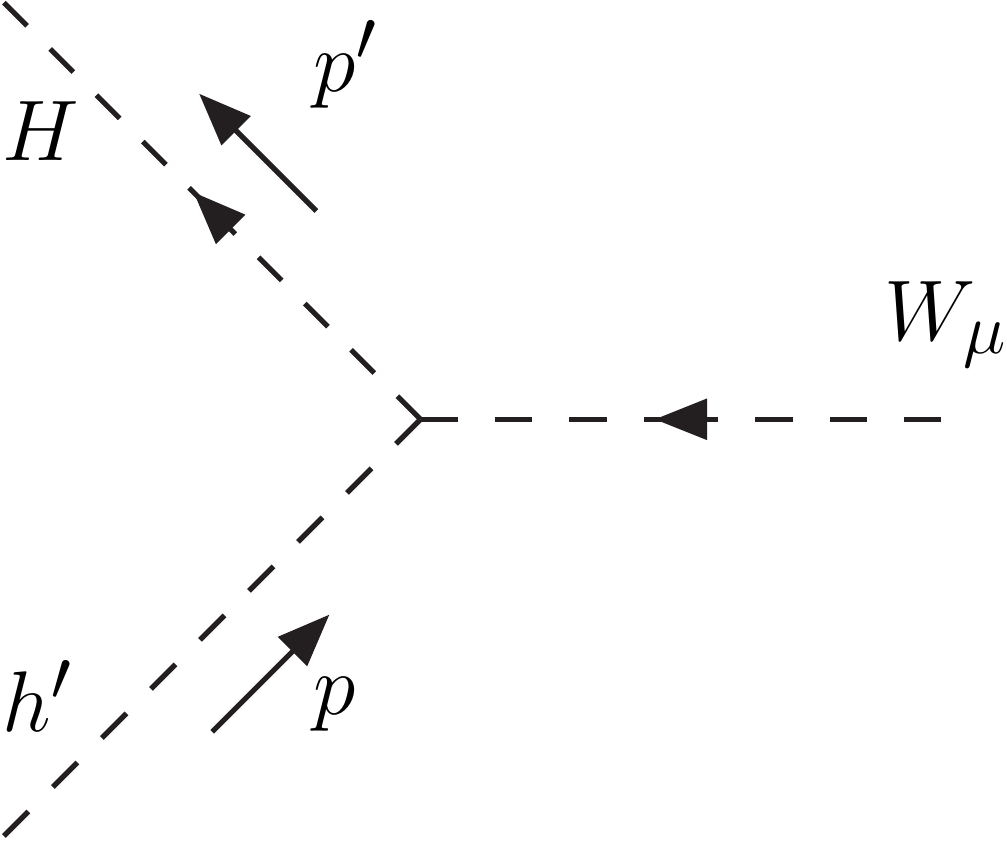} 
\hspace{0.1\linewidth}
&
{\large $ i \frac{e}{2 s_W}
	(p + p')_\mu
$}
\\
\end{tabular}\\
%
% h'' H W 
%
\begin{tabular}{m{0.5\linewidth}m{0.4\linewidth}}
\includegraphics[width=.7\linewidth]{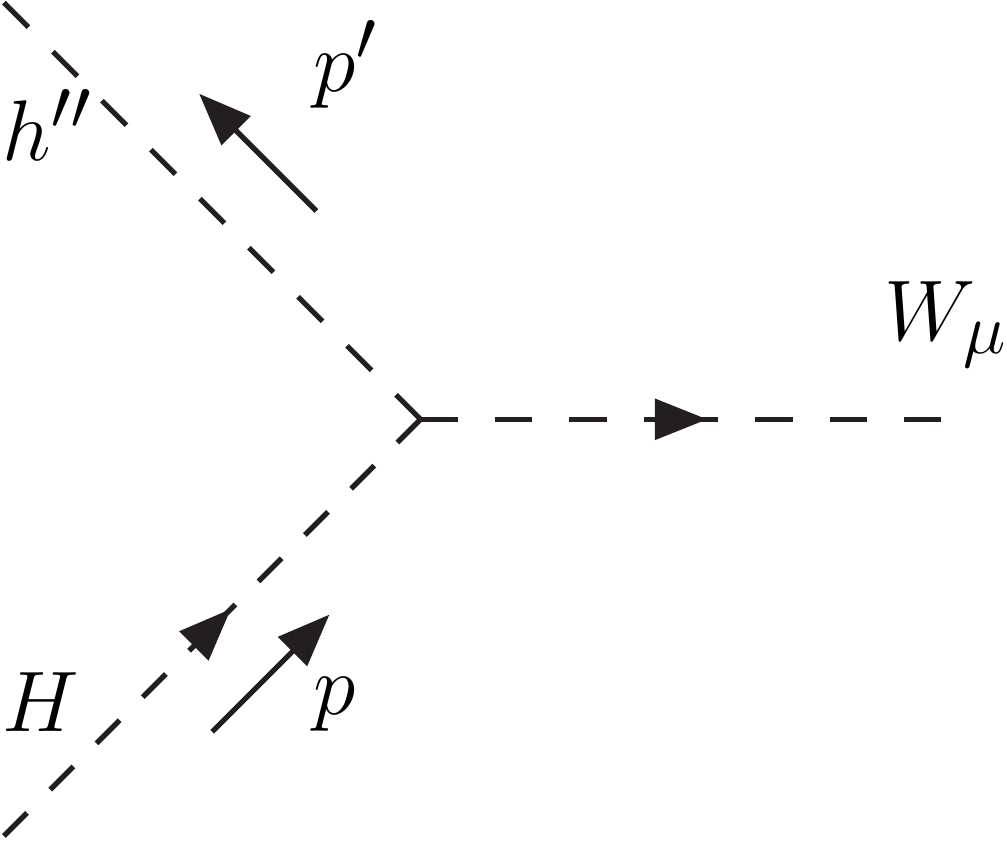} 
\hspace{0.1\linewidth}
&
{\large $ - \frac{e}{2 s_W}
	(p + p' )_\mu
$}
\\
\end{tabular}\\
%
% h'' H W 2
%
\begin{tabular}{m{0.5\linewidth}m{0.4\linewidth}}
\includegraphics[width=.7\linewidth]{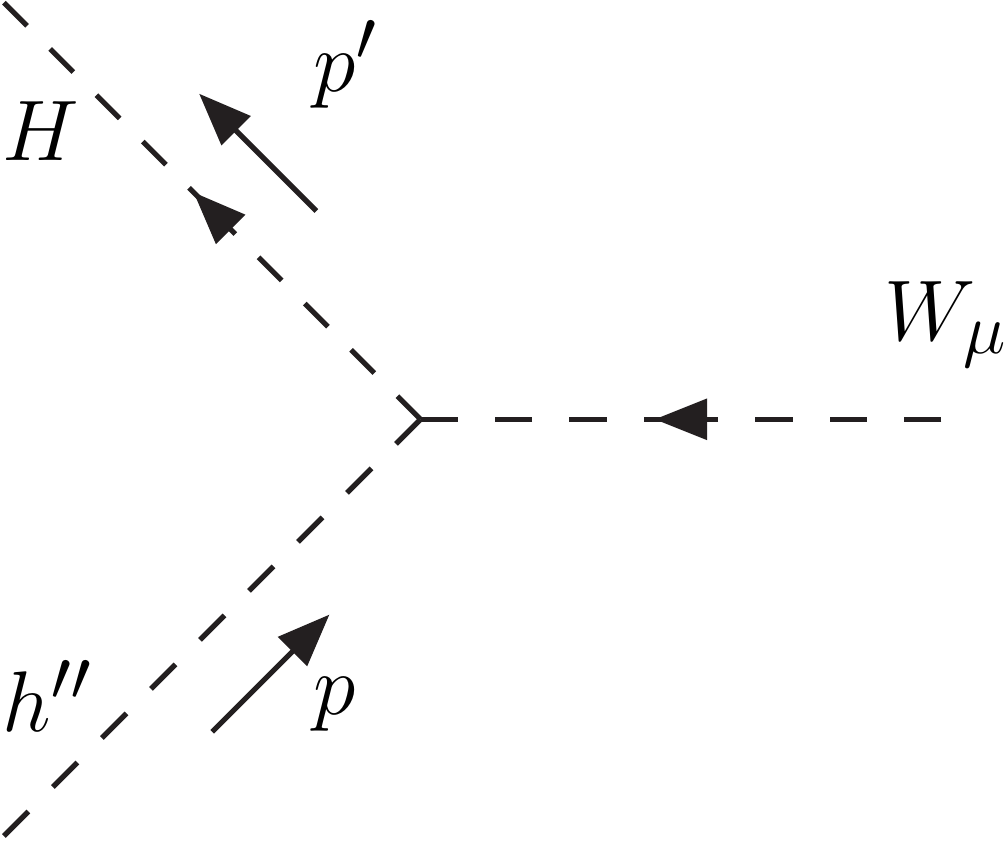} 
\hspace{0.1\linewidth}
&
{\large $ \frac{e}{2 s_W}
	(p + p' )_\mu
$}
\\
\end{tabular}\\

%%%%%%%%%%%%%%%%%%%%%%%%%%%%%%%%%%%%%%%%%%%%%%%%%%%%%%%%%%%%%%%%%%%%%%%
%
%
% REFERENCES
%

\end{document}